\documentclass[aps,pre,twocolumn,superscriptaddress,longbibliography,floatfix]{revtex4-1}

\usepackage{amsmath,amssymb,mathtools, bm,braket}
\usepackage{physics}      
\usepackage{siunitx}      

\PassOptionsToPackage{x11names}{xcolor}
\usepackage{graphicx}
\usepackage[caption=false]{subfig} 
\usepackage{tikz}
\usetikzlibrary{positioning}
\usepackage{microtype}
\usepackage{hyphenat}

\usepackage{times}     

\usepackage{xcolor}  
\usepackage[colorlinks=true,
            linkcolor=blue,
            citecolor=blue,
            urlcolor=blue,
            anchorcolor=red]{hyperref}

\begin{document}
	
	
	\title{Reservoir-Engineered Mechanical Cat States with a Driven Qubit}
	\author{M. Tahir Naseem} 
    \affiliation{Faculty of Engineering Science, Ghulam Ishaq Khan Institute of Engineering Sciences and Technology, Topi 23640, Khyber Pakhtunkhwa, Pakistan}
	\email{mnaseem16@ku.edu.tr}

    \begin{abstract}
Macroscopic quantum superpositions, such as mechanical Schrödinger-cat states, are central to emerging quantum technologies in sensing and bosonic error-correcting codes. We propose a scheme to generate such states by coupling a nanomechanical resonator to a coherently driven two-level system via both transverse and longitudinal interactions. Driving the qubit at twice the oscillator frequency activates resonant two-phonon exchange processes, enabling coherent conversion of drive energy into phonon pairs and their dissipative stabilization. Starting from the full time-dependent Hamiltonian, we derive an effective master equation for the mechanical mode by perturbative elimination of the lossy qubit. The reduced dynamics feature engineered two-phonon loss and a coherent squeezing term, which together drive the resonator into a deterministic Schrödinger-cat state. Our approach requires only a single driven qubit and no auxiliary cavity, offering a scalable and experimentally accessible route to macroscopic quantum superpositions in circuit-QED and related platforms.
    \end{abstract}
	\maketitle

\section{Introduction}

Quantum superposition lies at the heart of quantum mechanics and underpins Schr\"odinger-cat states—superpositions of macroscopically distinct configurations. Early breakthroughs in quantum optics realized such states in well-controlled microscopic platforms, including trapped ions and cavity fields~\cite{Monroe1996,Brune1996}. Extending cat-state preparation to massive mechanical resonators remains a central objective for probing the quantum–classical boundary and for enabling technologies in continuous-variable sensing and bosonic error-correcting codes~\cite{PhysRevLett.97.237201,Katz_2008,POOT2012273,Barzanjeh2022}. The challenge is significant: environmental decoherence grows rapidly with mass and size, demanding protocols that are both robust and experimentally economical. Realizing such states would provide new insights into macroscopic quantum behavior and enable applications in continuous-variable quantum information processing and quantum metrology~\cite{Woolley_2008,RevModPhys.85.623,RevModPhys.86.1391,pnas.1419326112}.

Rapid experimental progress has brought mechanical systems deep into the quantum regime, supported by foundational developments across multiple platforms~\cite{POOT2012273}. A key milestone was the ground-state cooling of a nanomechanical mode via coupling to a superconducting qubit~\cite{OConnell2010}, enabling single-phonon control and nonclassical motional states~\cite{PhysRevLett.121.123604,Wollack2022}. More broadly, hybrid platforms in which mechanical modes couple to two-level systems or electromagnetic cavities have enabled strong quantum control~\cite{RevModPhys.86.1391,pnas.1419326112,Clerk2020}, including the generation of high-order Fock states and qubit–mechanics entanglement~\cite{Chu2017,PhysRevLett.104.177205,Thomas2021}. In parallel, measurement-based schemes based on heralded photon or phonon detection have probabilistically produced non-Gaussian mechanical states~\cite{PhysRevA.98.011801,PhysRevLett.127.243601,PRXQuantum.2.030204}.
These advances are underpinned by extensive theoretical work on cat-state generation~\cite{PhysRevLett.58.1055,PhysRevLett.76.608,PhysRevA.55.3184,PhysRevA.81.042311,PhysRevA.86.013814,PhysRevA.93.033853,PhysRevA.98.063801, PhysRevA.100.013831, PhysRevA.99.013827, PhysRevA.99.022302, PhysRevA.101.043841, PhysRevLett.127.087203,PhysRevResearch.3.023088,PhysRevA.109.023703, PhysRevA.110.013711, PRXQuantum.6.010321}, and have led to experimental realization of cat-like states across a wide range of photonic and matter-based platforms~\cite{PhysRevA.45.5193,Monroe1996,PhysRevLett.97.083604,science.1122858,Wakui:07,PhysRevLett.101.233605,PhysRevA.82.031802,Gao2010,Yukawa:13,PhysRevLett.115.023602,Lo2015,Ulanov2016,Lewenstein2021,sciadv.abn1778,Yang2025,Yu2025,sciadv.adr4492}. Most recently, a 16-\textmu g bulk acoustic oscillator, containing approximately $10^{17}$ atoms, was conditionally prepared in a Schr\"odinger-cat state via qubit-mediated measurement back-action~\cite{science.adf7553}. Yet, achieving mechanical cat states that are both deterministic and long-lived, while avoiding post-selection or auxiliary cavities, remains an active area of research in macroscopic quantum control.

A complementary paradigm is quantum reservoir engineering, which targets deterministic preparation and stabilization of nonclassical states. In cavity and circuit QED systems, engineered dissipation has enabled the on-demand preparation of nonclassical states~\cite{PhysRevLett.57.13,PhysRevA.48.1582,Hofheinz2009,PhysRevA.87.042315,science.1243289,PhysRevA.110.043717}. A landmark demonstration by Leghtas \textit{et al.}\ realized two-photon dissipation in a superconducting resonator~\cite{Leghtas2015}, confining the dynamics to a subspace of coherent states with opposite phase and enabling the stabilization of a robust microwave cat state. This technique later enabled bias-preserving cat qubits with exponentially suppressed bit-flip errors~\cite{Ofek2016}. Extensions of this approach to mechanical systems have broadened the reach of reservoir engineering~\cite{PhysRevLett.110.253601,PhysRevA.105.012201, PhysRevLett.130.213604,PhysRevLett.134.043601}. Optomechanical cavities driven by two appropriately detuned tones can implement two-phonon cooling and driving processes that conserve parity~\cite{PhysRevA.104.063507,PhysRevLett.130.213604}. This mirrors the two-photon stabilization strategy of circuit QED and offers a path to robust macroscopic superpositions in long-lived mechanical systems.

Despite rapid progress, protocols for mechanical Schr\"odinger-cat states in solid-state resonators largely fall into three paradigms, each with distinct characteristics. (i) Measurement-based (heralded) schemes can achieve high fidelity but are intrinsically probabilistic and rely on post-selection~\cite{PhysRevA.101.033812,PRXQuantum.2.030204}. (ii) Cavity-mediated two-phonon dissipation, typically implemented with bichromatic cavity drives, enables deterministic steady-state stabilization but requires an auxiliary cavity and multi-tone control~\cite{PhysRevA.98.063801, PhysRevA.100.013831, PhysRevLett.130.213604}. (iii) Direct qubit--mechanics approaches (DC bias or single-tone conditional displacements) are generally transient and parameter-sensitive, offering no autonomous steady-state protection~\cite{Radić2022,PhysRevA.93.033853}. {\it{In this context, a minimal, cavity-free, single-tone architecture capable of deterministically stabilizing a parity-protected mechanical cat state is of particular interest—the goal of the scheme introduced next.}}

Here we propose a scheme for stabilizing mechanical Schr\"odinger\hyp{}cat states using a single driven qubit and no auxiliary cavity. A nanomechanical resonator couples to a coherently driven two-level system (qubit) via both transverse and longitudinal interactions. Tuning the qubit and drive near twice the mechanical frequency activates resonant two-phonon exchange: virtual pathways that combine the two couplings convert a qubit excitation coherently into a pair of phonons (and vice versa). Because the qubit relaxes rapidly, adiabatic elimination yields parity-preserving two-phonon dissipation for the resonator. Simultaneously, interference between the coherent drive and the nonlinear mixing generates a coherent two-phonon (squeezing) interaction. The coexistence of these ingredients drives the dynamics into the parity manifold determined by the oscillator’s initial parity, stabilizing an even- or odd-parity Schr\"odinger-cat steady state accordingly.
Starting from the full time-dependent Hamiltonian, we derive this reduced description by combining a perturbative unitary transformation that isolates the nonlinear interaction with a projection-operator elimination of the dissipative qubit~\cite{PhysRev.97.869,PhysRev.149.491}. The resulting effective master equation contains (i) qubit-induced single- and two-phonon dissipators and (ii) coherent two-phonon squeezing, along with qubit-mediated frequency shifts and an effective Kerr nonlinearity for the mechanical mode. This simple, single-tone architecture may provide a scalable, experimentally accessible route to macroscopic superpositions in circuit QED.

The microscopic origin of our two-phonon process is the same broken-inversion-symmetry mechanism exploited in Refs.~\cite{PhysRevA.99.022302,PhysRevA.110.013711}, where a driven, lossy flux qubit is used to generate Schr\"odinger-cat states of a microwave resonator or a magnon mode. However, our contribution differs both in scope and in the level of description. Here we list the key differences between our scheme and Refs.~\cite{PhysRevA.99.022302,PhysRevA.110.013711}: (i) We ask a different question: given a generic driven, dissipative two-level system with transverse and longitudinal couplings, what is the effective reduced Liouvillian for a mechanical oscillator alone once the qubit is adiabatically eliminated? (ii) At this Liouvillian level, the driven qubit and its environment realize an engineered two-phonon cooling reservoir that drives the mechanical resonator directly into a cat-state manifold  (cf. Appendix \ref{app:E}). The same two-phonon channel that stabilizes the superposition also cools the resonator below its bare thermal occupation; a cooling-into-a-cat-manifold mechanism that, to our knowledge, is not explicitly developed in Ref.~\cite{PhysRevA.99.022302}. (iii) The virtual processes that generate the two-phonon dissipation also induce a Kerr nonlinearity in the reduced Hamiltonian, which is known to generate cat state \cite{Puri2017}; however, this term is not explicitly present and hence does not play a central role in the analysis of Ref. \cite{PhysRevA.99.022302}. We derive this Kerr term microscopically from the qubit response functions, quantify its magnitude relative to the engineered two-phonon rate. We use this balance to formulate concrete design rules for shaping and stabilizing mechanical cat states, the details are presented in Appendix \ref{app:D} . (iv) Finally, in the same Appendix, we treat the mechanical resonator as a thermal, lossy mode and study how finite-temperature single-phonon damping competes with the engineered two-phonon reservoir in forming and stabilizing the cat state. We also compare sideband-resolved and unresolved regimes to derive design rules for the qubit linewidth, detuning, and couplings. In this sense, our work turns the Hamiltonian of Refs.~\cite{PhysRevA.99.022302,PhysRevA.110.013711} into a concrete reservoir-engineering toolkit for stabilizing mechanical cat states, rather than directly reusing their microwave-cavity scheme.

We present the driven qubit–resonator model in Sec.~\ref{sec:model}, derive the effective master equation and analyze the stabilization mechanism in Sec.~\ref{sec:results}, and conclude in Sec.~\ref{sec:conclusion}. Appendix~\ref{Appendix:A} discusses experimental feasibility, Appendix~\ref{Appendix:B} derives the reduced dynamics of the resonator, Appendix~\ref{Appendix:C} compares the full and effective models, Appendix~\ref{app:D} analyzes thermal decoherence, Kerr nonlinearity, and sideband resolution, and Appendix~\ref{app:E} connects the two-phonon cooling with cat state formation.

            \section{Physical Model}\label{sec:model}
We now introduce the microscopic model that forms the basis of our scheme. The system comprises a nanomechanical resonator coupled to a driven two-level system (qubit) through both longitudinal and transverse interactions. We describe the unitary dynamics, dissipation mechanisms, and the emergence of effective two-phonon processes responsible for stabilizing cat states in the mechanical mode.

        \subsection{System Hamiltonian}
      
We consider a nanomechanical resonator of frequency \( \omega_m \) coupled to a driven two-level system (qubit). The total Hamiltonian captures the free evolution of both subsystems, their mutual interactions, and a time-dependent external drive. Setting \( \hbar = k_B = 1 \) throughout, it takes the form~\cite{Niemczyk2010, PhysRevA.98.042331, PhysRevLett.117.043601, PhysRevResearch.4.013152}:
\begin{align}
H =\ & \omega_m\, a^\dagger a 
+ \frac{ \omega_q}{2} \sigma_z 
+ g_x\, \sigma_x (a + a^\dagger) 
+ g_z\, \sigma_z (a + a^\dagger) \nonumber \\
& + \Omega \cos(\omega_d t)\, \sigma_x,
\label{eq:SysHam}
\end{align}
where \( a \) and \( a^\dagger \) are the annihilation and creation operators for the mechanical mode, and \( \sigma_{x,z} \) are Pauli matrices acting on the qubit. The parameter \( \omega_q \) denotes the qubit transition frequency; \( g_x \) and \( g_z \) are the transverse and longitudinal coupling strengths, respectively; and \( \Omega \cos(\omega_d t) \) represents a classical drive applied along the qubit’s \( x \)-axis. Here, \( \Omega \) is the Rabi frequency, which characterizes the strength of coherent driving applied to the qubit. We emphasize that one need not engineer two independent couplings. A single capacitive (charge) coupling suffices: in the charge basis it is purely longitudinal, and after rotating to the qubit eigenbasis it decomposes into longitudinal and transverse components with relative weights set by the bias point (mixing angle)~\cite{Niemczyk2010, PhysRevA.98.042331, PhysRevA.99.022302, PhysRevA.110.013711}. A detailed discussion of the experimental feasibility of this model is provided in Appendix~\ref{Appendix:A}.

The dissipative dynamics of the driven qubit--resonator system can be described by a master equation in Lindblad form \cite{9780199213900.001.0001}. We account for qubit relaxation and thermal coupling of the mechanical resonator to its environment. The full master equation is
\begin{align}\label{eq:master}
\dot{\rho} = -i[H, \rho] 
+ \kappa\, \mathcal{D}[\sigma_-]\rho 
+ \gamma_-\, \mathcal{D}[a]\rho 
+ \gamma_+\, \mathcal{D}[a^\dagger]\rho,
\end{align}
where \( \sigma_- = |g\rangle\langle e| \) is the qubit lowering operator, and the Lindblad superoperator is defined as
\begin{align}
\mathcal{D}[L]\rho \equiv L \rho L^\dagger - \tfrac{1}{2} \left( L^\dagger L \rho + \rho L^\dagger L \right).
\end{align}
The term with rate \( \kappa \) accounts for qubit relaxation via spontaneous emission into its electromagnetic environment. The phonon loss and excitation rates are given by
\begin{align}
\gamma_- = \gamma (n_{\mathrm{th}} + 1), \quad \gamma_+ = \gamma n_{\mathrm{th}},
\end{align}
with \( n_{\mathrm{th}} = (e^{\omega_m/T} - 1)^{-1} \) the thermal phonon occupation at bath temperature \( T \), and \( \gamma \) the intrinsic mechanical damping rate. {\color{black}
In Eq.~\eqref{eq:master} we assume a dilution-refrigerator environment for the qubit, so that the thermal occupation of the qubit transition, $\bar n_q$, is effectively zero. In this limit, only the spontaneous-emission channel contributes appreciably, while the absorption channel is suppressed by the small factor $\bar n_q$. For typical circuit-QED parameters \cite{Blais2021}, with $\omega_q/2\pi$ in the range $5$--$10~\mathrm{GHz}$ and $T$ in the range $10$--$20~\mathrm{mK}$, one has $\hbar\omega_q/(k_B T)\gg 1$ and hence $\bar n_q \ll 1$. If a given device exhibits an elevated effective qubit temperature, the additional term $\kappa\,\bar n_q\,\mathcal{D}[\sigma_+]$ should be included in Eq.~\eqref{eq:master}. We note that the joint dynamics in our scheme correspond to a standard driven nonequilibrium open quantum system with distinct local reservoirs. Accordingly, Eq.~(\ref{eq:master}) describes a resonator coupled to a finite-temperature bath together with a qubit effectively coupled to a zero-temperature bath.
}

\subsection{Physical Interpretation and Two-Phonon Processes}

The dynamics of the qubit--resonator system are governed by two key couplings---transverse and longitudinal---as introduced in Eq. (\ref{eq:SysHam}). The transverse term enables direct energy exchange: a qubit flip is accompanied by the creation or annihilation of a single phonon. By contrast, the longitudinal term does not, by itself, exchange energy with the resonator; rather, it acts as a qubit-state--dependent force that shifts the resonator’s equilibrium position. When the qubit is driven, this force becomes time dependent and induces coherent displacements, a mechanism widely used to generate entangled qubit--mechanics states in closed, unitary evolutions~\cite{PhysRevB.109.155304}. In the present, strongly dissipative setting, the qubit can be adiabatically eliminated, and the combined action of the two couplings produces effective nonlinear processes within the resonator alone (see Appendix~\ref{Appendix:B}).

To selectively enhance two-phonon transitions, we operate near a two-phonon (second-order) resonance. Concretely, the qubit is driven close to its transition frequency (\(\omega_d \simeq \omega_q\)) while the qubit splitting is set near twice the mechanical frequency (\(\omega_q \simeq 2\omega_m\)). Under these conditions, virtual paths in which the transverse and longitudinal couplings act in sequence connect \(\ket{g,n}\) and \(\ket{e,n-2}\) with comparable energy denominators, producing an effective interaction of the form $
H_{\mathrm{eff}} \propto (a^{2}\sigma_{+} + a^{\dagger 2}\sigma_{-})$,
with strength that scales as \(g_{\mathrm{eff}} \sim g_x g_z/\omega_m\). Because the qubit relaxes rapidly, these virtual excitations generate qubit-mediated two-phonon channels for the resonator described by the dissipators \(\mathcal{D}[a^{2}]\) and \(\mathcal{D}[a^{\dagger 2}]\). Near \(\omega_q \approx 2\omega_m\) the two-phonon cooling rate \(\Gamma_{2}^{-}\) dominates over the heating rate \(\Gamma_{2}^{+}\), so the engineered dissipation removes energy in pairs and preserves phonon-number parity, in contrast to standard first-order sideband cooling schemes that target single-phonon processes at \(\omega_q - \omega_m\)~\cite{Jaehne_2008}. For clarity, we define the phonon-number parity operator \(P = \exp(i\pi a^\dagger a) = (-1)^{\hat{n}}\). Since \([P,a^2] = [P,a^{\dagger 2}] = 0\), the engineered two-phonon terms conserve parity, whereas the single-phonon processes flip parity. Related second-order, drive-assisted photon-pair processes have been demonstrated in superconducting-qubit platforms coupled to mechanical resonators~\cite{PhysRevB.97.125429}.

Cat-state stabilization in our scheme requires, in addition to two-phonon dissipation, a coherent two-phonon (squeezing) interaction. This element is known to drive the oscillator into superpositions of macroscopically distinct states and is a standard ingredient in cat-state engineering~\cite{PhysRevA.49.2785}. In our setup, interference between the coherent qubit drive and the nonlinear mixing generated by the combined transverse and longitudinal couplings yields an effective squeezing term \(\chi\, a^{2} + \chi^{*} a^{\dagger 2}\) in the reduced oscillator Hamiltonian. 
On resonance \((\omega_d = \omega_q)\), one has \(|\chi| \propto \Omega g_x g_z / (\kappa\omega_m)\) (Appendix~\ref{Appendix:B}), confirming that the squeezing term arises from the interplay between the drive and both transverse and longitudinal interactions. The coexistence of parity-preserving two-phonon cooling \((\Gamma_{2}^{-}\gg\Gamma_{2}^{+})\) and coherent two-phonon squeezing drives the dynamics into the parity sector determined by the initial resonator state. Consequently, the steady state is a Schr\"odinger-cat of definite parity: even if the initial state has even parity, and odd if
the initial state has odd parity. The Wigner function then exhibits the characteristic two-lobe structure with interference fringes. To summarize, we operate near a two-phonon sideband rather than the usual single-phonon resonance. Concretely, we tune the qubit close to twice the mechanical frequency and drive it near resonance (\(\omega_q \simeq 2\omega_m\), \(\omega_d \simeq \omega_q\)). This choice makes the qubit-mediated two-phonon exchange resonant, so that two-phonon cooling dominates over heating and single-phonon channels. While the interference between the drive and the nonlinear mixing produces a nearly stationary and sizable two-phonon squeezing term. In this regime, the engineered two-phonon dissipation and coherent squeezing act together to funnel the dynamics into a parity-protected Schr\"odinger-cat steady state.

    \section{Results}\label{sec:results}

In this section, we derive an effective master equation that governs the reduced dynamics of the mechanical resonator. To illustrate its physical implications, we compute the Wigner function to visualize the emergence of a Schrödinger-cat state. Starting from the full time-dependent Hamiltonian, we apply a unitary polaron-like transformation \cite{PhysRevB.57.347} to isolate nonlinear interactions between the qubit and the oscillator. We then perform an adiabatic elimination of the qubit using the Nakajima--Zwanzig formalism, valid in the regime of strong qubit dissipation. The resulting reduced dynamics incorporate both coherent and dissipative effects that stabilize a nonclassical steady state in the oscillator.

        \subsection{Adiabatic Elimination of the Qubit}

To derive a reduced master equation for the mechanical resonator, we adiabatically eliminate the qubit under the assumption that it is strongly dissipative and rapidly relaxes to its steady state. This procedure combines unitary perturbation theory with projection operator techniques. As a first step, we apply a unitary transformation to perturbatively eliminate the longitudinal coupling between the qubit and the resonator \cite{PhysRevB.57.347, Blais2021}. The longitudinal interaction in the original Hamiltonian [Eq.~(\ref{eq:SysHam})],
\begin{equation}
V = g_z \sigma_z (a + a^\dagger),
\end{equation}
introduces nonlinearities and breaks excitation-number conservation. To remove this term to leading order, we introduce an anti-Hermitian generator
\begin{equation}
S = -\frac{g_z}{\omega_m} \sigma_z (a^\dagger - a),
\end{equation}
and apply the unitary transformation \( H' = e^S H e^{-S} \). Expanding the result to leading order in the small parameter \( g_z / \omega_m \), the transformed Hamiltonian becomes
\begin{equation}\label{eq:SWT}
H' \approx H_0 + V + [S, H_0] + [S, V] + \frac{1}{2} [S, [S, H_0]],
\end{equation}
where \( H_0 \) contains all terms of the original Hamiltonian except the longitudinal coupling. Explicitly, we have
\begin{equation}
\begin{aligned}\label{eq:HamSW}
H_0 &= \frac{\omega_q}{2} \sigma_z + \omega_m a^\dagger a 
+ g_x \sigma_x (a + a^\dagger) \\
&\quad + \frac{\Omega}{2}(e^{i \omega_d t} + e^{-i \omega_d t}) \sigma_x.
\end{aligned}
\end{equation}
Evaluating the commutators in Eq.~(\ref{eq:SWT}) and retaining only the leading-order nonlinear contribution, we obtain the transformed Hamiltonian:
\begin{equation}\label{eq:SWHprime}
H' \approx H_0 - \frac{2i\, g_x g_z}{\omega_m} \sigma_y (a^{\dagger 2} - a^2).
\end{equation}
We note that our first step (Eq.~\eqref{eq:SWT}) is a purposeful unitary change of frame rather than a textbook Schrieffer--Wolff construction~\cite{PhysRev.149.491} that cancels all couplings. We design this step deliberately to displace only the longitudinal force while retaining the transverse vertex. This choice leaves the single-phonon spin-flip coupling in place so that, together with the displaced $\sigma_z$ force, the leading cross-commutator generates the two-phonon operator at order $g_x g_z/\omega_m$. This term is resonant near $\omega_q \simeq 2\omega_m$ and is precisely the pathway that produces the engineered two-phonon dissipation and the coherent squeezing in the reduced dynamics. Operationally, the same displacement also simplifies the subsequent projection-operator elimination. After the transformation, the interaction has zero qubit mean,  so the first-order Nakajima--Zwanzig term vanishes (see Appendix~\ref{Appendix:B} for details). We can then choose a qubit-only, time-independent reference Liouvillian, \(\mathcal{L}_0 = \kappa\,\mathcal{D}[\sigma_-]\), for which \(e^{\mathcal{L}_0 \tau}\sigma_\pm = e^{-\kappa \tau/2}\sigma_\pm\). As a result, the second-order memory kernels collapse to simple scalar response functions, directly yielding the compact rates together with the Lamb and Kerr shifts. In this precise Liouvillian sense, the unitary transformation ``simplifies the dynamics'' rather than removing all interactions or time dependences.
The first term \( H_0 \) in Eq. \eqref{eq:SWHprime} describes the unperturbed evolution of the system, while the second term, generated by the transformation, captures a nonlinear two-phonon interaction. This effective coupling arises from the combined action of the longitudinal and transverse qubit--resonator interactions and enables simultaneous creation or annihilation of phonon pairs. Such two-phonon terms are central to the resonant dynamics responsible for cat-state generation. 
The dynamics can be simplified and resonant terms isolated by transitioning into a rotating frame defined by the unitary transformation
\begin{equation}
    U(t) = \exp(-i \omega_q t \, \sigma_z/2) \exp(-i \omega_m t \, a^\dagger a).
\end{equation}
Applying this transformation to Eq.~(\ref{eq:HamSW}), we obtain the time-dependent interaction-picture Hamiltonian:
\begin{align}\label{eq:Hint}
H_{\text{int}} &=  g_x (\sigma_+ e^{i \omega_q t}  + \sigma_-e^{-i \omega_q t} )
( a\, e^{-i \omega_m t} + a^\dagger\, e^{i \omega_m t} )  \nonumber \\& 
+  g (  \sigma_+ e^{+i \omega_q t}  - \sigma_- e^{-i \omega_q t}  )
( a^{\dagger 2} e^{2 i \omega_m t} - a^2 e^{-2 i \omega_m t} ) \nonumber \\
& +  \varepsilon ( \sigma_+ e^{i \omega_q t}  + \sigma_- e^{-i \omega_q t} )( e^{i \omega_d t} + e^{-i \omega_d t} ),
\end{align}
where \( g =  g_x g_z / \omega_m \) and \( \varepsilon = \Omega/2 \). 

To derive an effective reduced description for the mechanical resonator, we begin by expressing the dynamics of the coupled qubit--resonator system in terms of Liouvillian superoperators. Our goal is to eliminate the qubit degrees of freedom, assuming it is strongly dissipative and rapidly relaxes to its steady state 
In this regime, the qubit acts as a lossy auxiliary subsystem that mediates effective nonlinear processes in the mechanical mode.
We employ the Nakajima--Zwanzig projection operator formalism to systematically obtain the reduced dynamics of the resonator~\cite{10.1143/PTP.20.948, 10.1063/1.1731409, ZWANZIG19641109}, following approaches used in earlier studies of qubit-assisted cooling~\cite{Jaehne_2008}. The full evolution of the density matrix \( \rho \) is governed by the Liouvillian equation
\begin{equation}
\frac{d\rho}{dt} = \mathcal{L} \rho = (\mathcal{L}_0 + \mathcal{V}) \rho,
\end{equation}
where \( \mathcal{L}_0 \) captures qubit dynamics: dissipation through spontaneous emission, given by
\begin{equation}
\mathcal{L}_0 \rho = \kappa\, \mathcal{D}[\sigma_-] \rho.
\end{equation}
By contrast, \( \mathcal{V} \) contains the time-dependent qubit--resonator coupling as well as the thermal damping of the mechanical oscillator:
\begin{equation}
\mathcal{V} \rho = -i[H_{\text{int}}(t), \rho] + \gamma_- \mathcal{D}[a]\rho + \gamma_+ \mathcal{D}[a^\dagger]\rho.
\end{equation}
Here, \( H_{\text{int}}(t) \) is the interaction Hamiltonian in the rotating frame, as defined in Eq.~(\ref{eq:Hint}).

We extract the slow dynamics of the mechanical subsystem by introducing a projection superoperator \( \mathcal{P} \), which maps the full system state onto the tensor product of the reduced resonator state and the steady state of the qubit. The projection is defined as
\begin{equation}
\mathcal{P} \rho(t) =   \rho_\text{q}^{\text{ss}} \otimes \text{Tr}_\text{q}[\rho(t)],
\end{equation}
where \( \text{Tr}_\text{q} \) denotes the partial trace over the qubit degrees of freedom, yielding the reduced density matrix of the mechanical resonator. The state \( \rho_\text{q}^{\text{ss}} \) is the stationary state of the qubit. In the present analysis, we assume the qubit is coupled to a thermal reservoir at zero temperature, so that its steady state is the ground state: $\rho_\text{q}^{\text{ss}} = |g\rangle\langle g|$.
More generally, however, one could instead choose \( \rho_\text{q}^{\text{ss}} \) to be a thermal state if the qubit were coupled to a finite-temperature bath.  The complementary projector \( \mathcal{Q} = 1 - \mathcal{P} \) captures all fast qubit-induced transients.
Applying the Nakajima--Zwanzig formalism and taking the standard Born–Markov limit in the fast qubit subspace yields the second‑order memory‑kernel equation \cite{Jaehne_2008}:
\begin{equation}\label{eq:Prho}
\frac{d}{dt}\mathcal{P}\rho(t) = \mathcal{P}\mathcal{L}\mathcal{P}\rho(t) + \int_0^\infty d\tau\, \mathcal{P} \mathcal{L}(t) e^{\mathcal{Q} \mathcal{L}_0\tau} \mathcal{Q} \mathcal{L}(t - \tau) \mathcal{P} \rho(t).
\end{equation}
The first term describes direct evolution within the projected subspace, while the second encodes second-order memory effects arising from virtual qubit excitations. Evaluating this memory kernel yields an effective master equation for the resonator alone (see Appendix~\ref{Appendix:B}):
\begin{align}\label{eq:EffME}
\frac{d\rho_\text{m}}{dt} &= -i[H_\text{eff}, \rho_\text{m}] + (\Gamma_1^- + \gamma_-)\mathcal{D}[a] \rho_\text{m} \nonumber \\
&
+ (\Gamma_1^+ + \gamma_+)\mathcal{D}[a^\dagger] \rho_\text{m}  + \Gamma_2^- \mathcal{D}[a^2] \rho_\text{m} + \Gamma_2^+ \mathcal{D}[a^{\dagger 2}] \rho_\text{m}.
\end{align}
Here, the two-phonon dissipators are derived from a second-order Nakajima–Zwanzig elimination of the driven, lossy qubit; they are not included phenomenologically, see Sec. \ref{subsec:M_22} in Appendix \ref{Appendix:B} for details. The effective Hamiltonian governing the coherent evolution of the mechanical mode reads
\begin{equation}
H_\text{eff} = \omega_m^\text{eff} a^\dagger a + \chi a^2 + \chi^* a^{\dagger 2} + \delta_k (a^\dagger a)^2,
\end{equation}
where the renormalized frequency \( \omega_m^\text{eff} = \omega_m + \delta_1 + \delta_2 \) includes contributions from qubit-induced energy shifts.  The effective Hamiltonian $H_\text{eff}$ follows from the coherent parts of the second-order kernels \(M_{11}\), \(M_{22}\), and \(M_{23}\) derived in Appendix \ref{Appendix:B}, which produce the coefficients \(\delta_{1}\), \(\delta_{2}\), \(\delta_{k}\), and \(\chi\).
The coefficient \( \delta_1 \) represents the single-phonon Lamb shift arising from linear coupling between the qubit and resonator, while \( \delta_2 \) originates from virtual two-phonon exchange processes involving the nonlinear interaction. In contrast, \( \delta_k \) corresponds to an effective Kerr nonlinearity for the resonator mode, generated by higher-order virtual transitions through the qubit. 
These frequency shifts are given explicitly by
\begin{align}
\delta_1 &= g_x^2 \left[ \text{Im}(S_-) + \text{Im}(S_+) \right], \nonumber \\
\delta_2 &= g^2 \left[ -\text{Im}(S_{2-}) + 3\,\text{Im}(S_{2+}) \right], \nonumber \\
\delta_k &= g^2 \left[ \text{Im}(S_{2-}) + \text{Im}(S_{2+}) \right],
\label{eq:delta_k}
\end{align}
and the corresponding dissipative rates induced by the eliminated qubit are
\begin{equation}
\Gamma_1^\pm = 2\, g_x^2\, \text{Re}(S_\pm), \quad \Gamma_2^\pm = 2\,g^2\, \text{Re}(S_{2\pm}),
\label{eq:Gamma_pm}
\end{equation}
with the qubit response functions defined as
\begin{equation}
S_\pm = \frac{1}{\kappa/2 + i(\omega_q \pm \omega_m)}, \quad
S_{2\pm} = \frac{1}{\kappa/2 + i(\omega_q \pm 2\omega_m)}.
\label{eq:S_pm}
\end{equation}
The coherent two-phonon interaction strength \( \chi \), which arises from the interplay between the transverse drive and nonlinear qubit–resonator coupling, takes the form
\begin{equation}
\chi = -\frac{i \varepsilon g}{\kappa/2 - i(\omega_q - \omega_d)}.
\label{eq:chi}
\end{equation}
When the condition \( \omega_q = \omega_d  \) is satisfied, this two-phonon squeezing term becomes resonant. Simultaneously, the two-phonon cooling rate \( \Gamma_2^- \) is enhanced, while the heating rate \( \Gamma_2^+ \) is strongly suppressed. This combination of engineered dissipation and coherent nonlinear driving enables the steady-state generation of Schrödinger-cat states in the mechanical resonator. 
\begin{figure*}[t]
    \centering
    \includegraphics[width=0.24\textwidth]{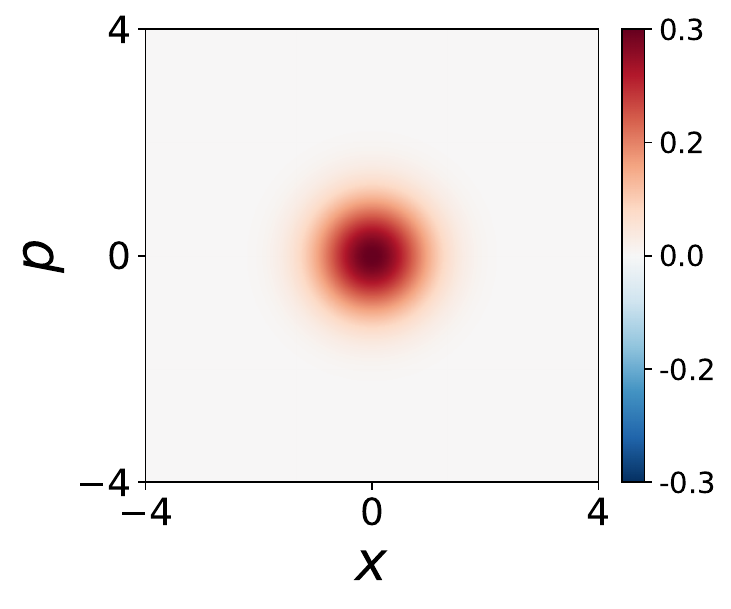}
    \includegraphics[width=0.24\textwidth]{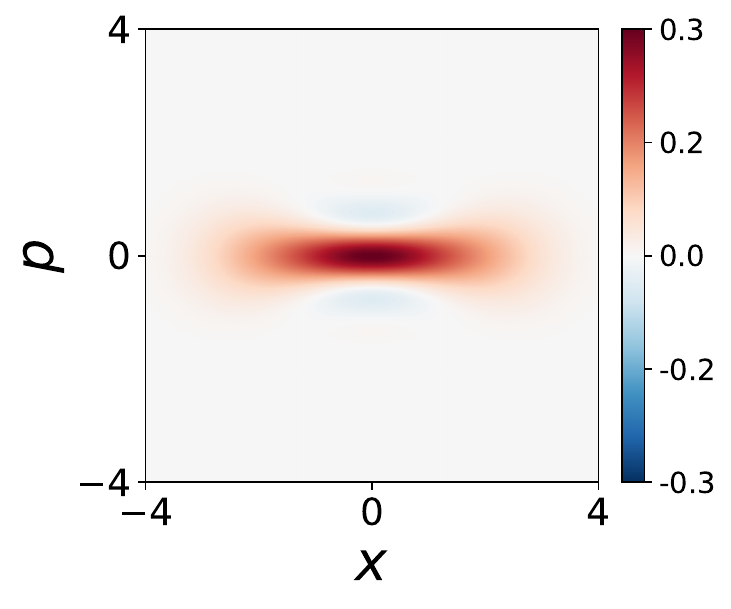}
    \includegraphics[width=0.24\textwidth]{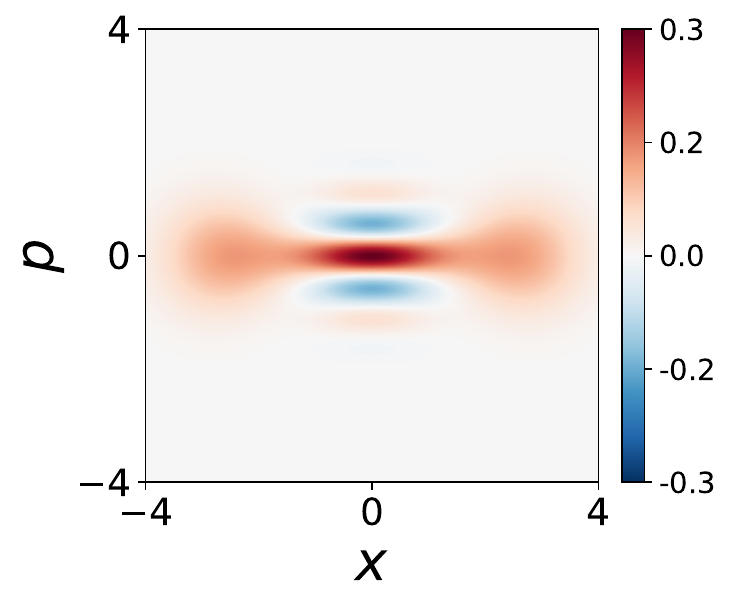}
    \includegraphics[width=0.24\textwidth]{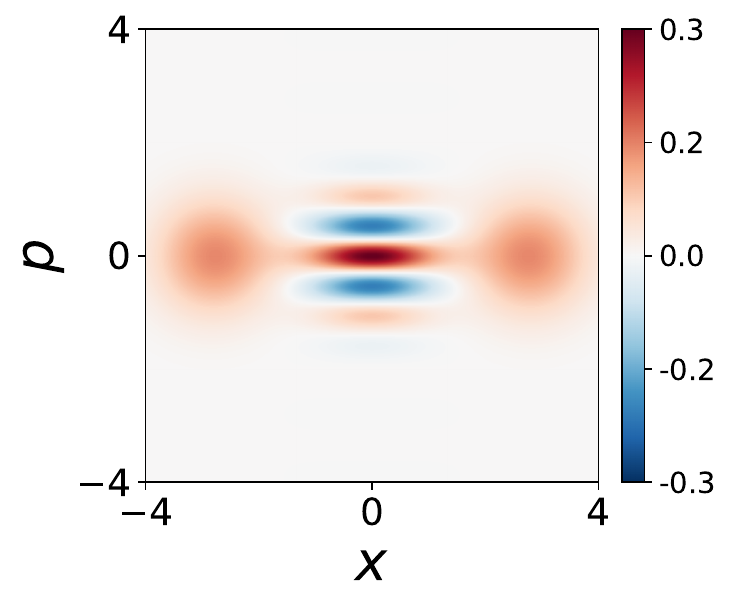}
    \caption{Time evolution of the mechanical resonator's state under the effective master equation (\ref{eq:EffME}), visualized via the Wigner quasi-probability distribution. The panels from left to right represent the Wigner function of the mechanical resonator at $\Gamma^{-}_{2} t = 0, 0.3, 0.6, 1$. The resonator starts in the vacuum state, appearing as a single peak at the origin in phase space at $t=0$. 
     Subsequently, the Wigner distribution bifurcates into two lobes, indicating the formation of a superposition of two coherent-state components. By the steady state (final frame), an even Schr\"odinger-cat state is realized, featuring two distinct phase-space peaks and a set of interference fringes between them. Parameters:  \( g_z/2\pi =  6~\text{MHz} \), \( g_x = 0.1 g_z \),  \( \omega_m/2\pi = 100~\text{MHz} \), \( \kappa/2\pi =   100~\text{kHz} \), \( \gamma/2\pi =  15~\text{Hz} \), \( \varepsilon =  4g \). We set the qubit and drive frequencies resonantly at \( \omega_q = \omega_d = 2\omega_m \) to activate the two-phonon dynamics and assume a zero-temperature environment for the mechanical bath. The mechanical Hilbert space is truncated to 60.}
    \label{fig:fig1}
\end{figure*}

        \subsection{Cat-State Generation}

To visualize the nonclassical features of the mechanical resonator, we compute the Wigner quasi-probability distribution \(W(x,p)\) of its quantum state. The Wigner function is a phase-space representation that captures quantum coherence via negative and oscillatory regions with no classical analogue~\cite{PhysRev.40.749}. It is defined by
\begin{equation}
W(x,p) = \frac{1}{\pi} \int_{-\infty}^{\infty} \mathrm{d}y\, \langle x - y \,|\, \rho \,|\, x + y \rangle\, e^{2 i p y},
\end{equation}
where \(\rho\) is the resonator density matrix and \(x,p\) are dimensionless quadratures. Interference fringes in \(W(x,p)\) are a hallmark of Schr\"odinger-cat superpositions.

We simulate the mechanical dynamics under the effective master equation~(\ref{eq:EffME}) to track how these features emerge in time. We adopt experimentally accessible parameters~\cite{RevModPhys.85.623, OConnell2010, Niemczyk2010, PhysRevLett.102.090501, science.aao1511}: \(g_z/2\pi = 6~\mathrm{MHz}\), \(g_x = 0.1\,g_z\); mechanical frequency \(\omega_m/2\pi = 100~\mathrm{MHz}\); qubit decay rate \(\kappa/2\pi = 100~\mathrm{kHz}\); and mechanical damping \(\gamma/2\pi = 15~\mathrm{Hz}\). We set the qubit and drive resonantly to \(\omega_q = \omega_d = 2\omega_m\) to activate two-phonon processes, and we take a zero-temperature mechanical bath in the numerics (i.e., \(n_{\mathrm{th}}=0\)). The mechanical Hilbert space is truncated to 60 to ensure convergence. Fig.~\ref{fig:fig1} displays snapshots of \(W(x,p)\): starting from the vacuum (single Gaussian peak), the distribution splits into two lobes and develops interference fringes at intermediate times, and by \(\Gamma_2^- t \simeq 1\) it approaches a steady even-parity cat state with two well-separated peaks. A detailed comparison between the full master equation [Eq. (\ref{eq:master})] and the reduced effective model [Eq. (\ref{eq:EffME})] is presented in Appendix~\ref{Appendix:C}. An approximate analytical expression for the steady–state mean phonon number in the cat–state regime, based on a two–phonon cooling picture, is derived in Appendix~\ref{app:E}.

The stabilization mechanism combines engineered two-phonon dissipation with coherent two-phonon (squeezing) dynamics, both mediated by the driven, lossy qubit. Near the two-phonon resonance \(\Delta_{2-} \equiv \omega_q - 2\omega_m \simeq 0\), second-order paths in which the transverse and longitudinal couplings act in sequence connect \(|g, n\rangle\) with \(|e, n-2\rangle\). These paths coherently sum to an effective operator \(a^{2}\sigma_{+} + a^{\dagger 2}\sigma_{-}\) with strength \(g \). Because the qubit relaxes rapidly, tracing it out yields qubit-induced two-phonon channels for the resonator. On resonance, the cooling channel dominates with \(\Gamma_2^- \gg \Gamma_2^+\) for \(\omega_q \simeq 2\omega_m\). The dissipator \(\mathcal{D}[a^2]\) removes energy in pairs and preserves phonon-number parity; consequently, when single-phonon channels (\(\Gamma_1^\pm,\gamma_\pm\)) are subdominant, the dynamics are funneled into the even-parity manifold (cf.\ the single-phonon sideband-cooling contrast in Ref.~\cite{Jaehne_2008}).

In parallel, interference between the coherent drive and the nonlinear mixing generated by the combined couplings produces a coherent two-phonon squeezing term in the reduced Hamiltonian, so that on resonance \(\omega_d = \omega_q\) one has \(|\chi| = 2\varepsilon g / \kappa\). The coexistence of strong, parity-preserving two-phonon cooling (\(\Gamma_2^- \gg \gamma\)) and coherent two-phonon squeezing produces a steady state with robust interference fringes and a characteristic two-lobe structure in the Wigner function. This behavior appears over a broad but specific hierarchy: \(g_{x,z} \ll \omega_m\) (validity of the polaron-like transformation given step in Eq. (\ref{eq:SWT})), \(\kappa \gg \{\Gamma_2^-,\,\Gamma_1^\pm,\,\gamma\}\) (adiabatic elimination), and \(\omega_q \simeq \omega_d \simeq 2\omega_m\) (two-phonon resonance). Related second-order, drive-assisted phonon-pair processes have been demonstrated in superconducting-qubit platforms~\cite{PhysRevB.97.125429}, consistent with the operating regime considered here. The impact of a finite–temperature phonon bath, the induced Kerr nonlinearity, and the degree of sideband resolution on the cat state dynamics is analyzed in detail in Appendix~\ref{app:D}.

\section{Conclusion}\label{sec:conclusion}

We have introduced a reservoir-engineering scheme for preparing Schr\"odinger-cat states in a nanomechanical resonator by coupling it to a coherently driven, dissipative two-level system. The core mechanism relies on exploiting both longitudinal and transverse couplings between the qubit and the mechanical mode while driving the qubit near twice the oscillator frequency. This configuration activates resonant two-phonon exchange processes, enabling energy from the drive to be coherently converted into phonon pairs within the resonator.
To describe the resulting dynamics, we started from a microscopic time-dependent Hamiltonian and applied a sequence of transformations. First, we used a unitary transformation to capture the effective nonlinear interactions. Then, by adiabatically eliminating the strongly dissipative qubit using the Nakajima--Zwanzig formalism, we derived an effective master equation for the mechanical resonator. The reduced equation features (i) qubit-induced single- and two-phonon dissipators with rates determined by qubit response functions, (ii) a coherent two-phonon squeezing term with amplitude \(\chi\) set by the drive--qubit detuning and linewidth, and (iii) qubit-mediated frequency shifts together with an effective Kerr nonlinearity. Their combined action drives the oscillator toward a nonclassical steady state characterized by well-defined parity and quantum interference in the Wigner function—hallmarks of a Schr\"odinger-cat state.

Our scheme contrasts with previous approaches that rely on conditional measurements, multi-tone driving, or active feedback~\cite{PhysRevA.93.033853,PhysRevA.98.063801, PhysRevA.100.013831, PhysRevLett.130.213604,PhysRevA.101.033812,Radić2022}. It enables deterministic cat-state generation using a single coherent drive, without requiring post-selection or monitoring. The simplicity and autonomy of the protocol make it particularly attractive for near-term implementation in hybrid quantum platforms such as superconducting circuits~\cite{RevModPhys.85.623} and spin--mechanical devices~\cite{PhysRevLett.113.020503,10.1063/5.0024001}.
Finally, the framework might be extended to multiple mechanical resonators coupled to a common driven qubit. In such systems, the qubit may mediate two-mode squeezing and correlated two-phonon dissipation, enabling nonclassical correlations between spatially separated modes and the stabilization of entangled cat states of the form \(\psi \propto \ket{\alpha,\alpha}+\ket{-\alpha,-\alpha}\).
More broadly, the interplay of engineered dissipation and coherent pair processes highlighted here suggests practical routes to mechanical bosonic encodings and quantum-enhanced sensing with macroscopic superpositions.

    \vspace{1em}
    \noindent\textbf{Data availability statement} \\
     The numerical simulation codes used in this study are available at ~\cite{CodeRepo}.
     
\appendix

\begin{widetext}


    \section{Experimental Platform and Feasibility}\label{Appendix:A}

To implement our scheme for the dissipative generation of mechanical Schrödinger-cat states, we require a controllable quantum platform that supports both longitudinal and transverse qubit–resonator interactions. A superconducting charge qubit such as a Cooper-pair box (CPB) or transmon or fluxonium, coupled to a nanomechanical resonator provides such an architecture, with both coupling types realizable through careful circuit design. Coherent control is enabled via external drives, while relaxation and decoherence arise from natural or engineered couplings to electromagnetic and phononic baths.
Here, we review and reconstruct the system Hamiltonian and master equation relevant to our proposed scheme. Our goal is to show how the model used in Eq.~(\ref{eq:master}) of the main text can be realized within a well-established superconducting qubit--mechanical resonator architecture \cite{PhysRevLett.88.148301, RevModPhys.85.623}. Simultaneous longitudinal and transverse coupling between superconducting qubits and mechanical resonators has been proposed theoretically and considered feasible within circuit QED architectures in several previous works (e.g.,~\cite{Niemczyk2010, PhysRevA.98.042331}). For the CPB, the longitudinal and transverse couplings both arise from the same capacitive pathway after rotation to the qubit eigenbasis; no extra circuit element is required beyond choosing the bias point~\cite{RevModPhys.73.357, Niemczyk2010, PhysRevA.98.042331, PhysRevA.99.022302, PhysRevA.110.013711, Blais2021}. Here, for completeness and clarity, we now describe the physical origin of each term in the Hamiltonian and explain the system’s coupling to environmental degrees of freedom. Building on these foundations, we further show how dissipation mechanisms can be incorporated in a consistent manner to yield the master equation~(\ref{eq:master}).

We consider a CPB capacitively coupled to a nanomechanical resonator. The CPB consists of a superconducting island connected to a reservoir via a Josephson junction with Josephson energy \(E_J\) and charging energy \(E_C\). The island is biased by a gate voltage \(V_g\) through a capacitance \(C_g(x)\) that depends on the resonator displacement \(x\); to leading order, \(C_g(x)\simeq C_{g0}+(\partial C_g/\partial x)\,x\). The qubit can be driven via gate-voltage modulation (tuning the gate charge \(n_g\)); if implemented as a split junction (dc SQUID), the Josephson energy \(E_J(\Phi)\) can also be modulated by magnetic flux. The CPB couples to an electromagnetic bath, while the mechanical mode interacts with a thermal phonon bath.
The Hamiltonian of the CPB in the charge basis is \cite{RevModPhys.73.357,devoret2004,You2011}
\begin{equation}
H_{\text{CPB}} = 4E_C (\hat{n} - n_g)^2 - E_J \cos \hat{\phi},
\end{equation}
where \(\hat{n}\) is the Cooper-pair number operator on the island and \(n_g = C_g V_g/2e\) is the dimensionless gate charge. Near the charge degeneracy point and within the two-level truncation \(\{|0\rangle,|1\rangle\}\), the qubit Hamiltonian reduces to \cite{RevModPhys.73.357,devoret2004,Niemczyk2010, You2011}
\begin{equation}
H_q^{(\text{charge})} = \frac{\epsilon}{2}\,\sigma_z - \frac{E_J}{2}\,\sigma_x, 
\qquad \epsilon = 4E_C\bigl(1-2 n_{g0}\bigr),
\end{equation}
where \(n_{g0}\) is the static gate bias. This Hamiltonian is diagonalized by a rotation \(U=\exp(-i \theta_0 \sigma_y/2)\), with
\begin{equation}
\tan\theta_0 = \frac{E_J}{\epsilon},\qquad 
\cos\theta_0 = \frac{\epsilon}{\omega_q},\qquad 
\sin\theta_0 = \frac{E_J}{\omega_q},\qquad
\omega_q = \sqrt{E_J^2+\epsilon^2}.
\end{equation}
In the qubit eigenbasis one has \(U H_q^{(\text{charge})} U^\dagger = (\omega_q/2)\sigma_z\), and the charge-basis Pauli operators transform as
\begin{equation}
\sigma_z^{(\text{charge})} = \sigma_z \cos\theta_0 + \sigma_x \sin\theta_0,\qquad
\sigma_x^{(\text{charge})} = -\,\sigma_z \sin\theta_0 + \sigma_x \cos\theta_0.
\end{equation}
The nanomechanical resonator (frequency $\omega_m$) has $H_m=\omega_m a^\dagger a$ and displacement $x = x_{\text{zpf}}(a+a^\dagger)$ with 
$x_{\text{zpf}}=\sqrt{1/(2 m \omega_m)}$. Its motion modulates the gate capacitance, which to leading order we expand as 
$C_g(x)\simeq C_g^{(0)}+ (dC_g/dx)\, x$, hence the gate charge
\begin{equation}
n_g(x) \simeq n_{g0} + \delta n_g, 
\qquad 
\delta n_g = \lambda\,(a+a^\dagger),
\qquad
\lambda = \frac{V_g}{2e}\frac{dC_g}{dx}\,x_{\text{zpf}}
\end{equation}
where $\lambda$ is a dimensionless coupling constant. Expanding the charging term to linear order in $\delta n_g$ gives
\begin{equation}
H_{\text{int}} \simeq -8E_C(\hat n-n_{g0})\,\delta n_g = 
-4E_C\,\lambda\,(a+a^\dagger)\,\sigma_z^{(\text{charge})},
\end{equation}
where in the two-level subspace we use $\hat n-n_{g0}\approx \sigma_z^{(\text{charge})}/2$. Transforming to the qubit eigenbasis using 
$\sigma_z^{(\text{charge})}=\sigma_z\cos\theta_0+\sigma_x\sin\theta_0$
yields the generic longitudinal and transverse electromechanical couplings
\begin{equation}
H_{\text{int}} = g_z\,\sigma_z (a+a^\dagger) + g_x\,\sigma_x (a+a^\dagger),
\qquad
g_z = -\,4E_C \lambda \cos\theta_0,\quad
g_x = -\,4E_C \lambda \sin\theta_0.
\end{equation}
This generalized coupling structure, including both longitudinal and transverse interactions, plays a crucial role in enabling nonlinear effects necessary for engineered dissipation and cat-state generation \cite{PhysRevA.99.022302, PhysRevA.110.013711, PhysRevLett.130.213604}. The coherent drive term introduced in Eq. (\ref{eq:SysHam}) can be implemented experimentally using an external microwave field, as discussed in prior studies \cite{PhysRevA.99.022302, PhysRevA.110.013711, Radić2022, PhysRevApplied.17.064022}.
While we work with a CPB here for concreteness, the same Hamiltonian structure and adiabatic-elimination procedure apply to a fluxonium qubit capacitively or piezoelectrically coupled to a nanomechanical mode~\cite{Manucharyan2009,Blais2021,PhysRevX.9.041041}. Fluxonium supports MHz–GHz-range transition frequencies with large anharmonicity, which makes the two-phonon resonance condition $\omega_q \approx 2\omega_m$ accessible for $\omega_m\!\sim\!10^2\text{--}10^3$\,\text{MHz}. Moreover, dispersive coupling between superconducting qubits and mechanical oscillators has been demonstrated experimentally, and standard Purcell engineering of $\kappa$ is available in circuit QED~\cite{Blais2021}.%

To complete the description of our system dynamics, we now account for dissipation mechanisms arising from interactions with the environment.
Dissipation in our setup primarily stems from two sources: electromagnetic noise coupled to the CPB and thermal phonon noise acting on the mechanical resonator.
We model gate-voltage fluctuations as a bath-driven charge offset
\(\delta n_g(t)=\sum_j \eta_j\,(c_j + c_j^\dagger)\),
so that expanding the charging energy to linear order in \(\delta n_g(t)\) yields the qubit–bath interaction~\cite{RevModPhys.73.357}
\begin{equation}
H_{I,q} = -8E_C\,(\hat n - n_{g0})\,\delta n_g(t).
\end{equation}
The interaction in the qubit eigenbasis is given by
\begin{equation}
H_{I,q} = -4E_C\!\left(\sigma_z \cos\theta_0 \;+\; \sigma_x \sin\theta_0\right)\!
\sum_j \eta_j\,(c_j + c_j^\dagger),
\end{equation}
revealing that both longitudinal and transverse noise components contribute: the \(\sigma_x\) channel induces relaxation, while the \(\sigma_z\) channel leads to pure dephasing.
The mechanical resonator, in contrast, is coupled to a thermal phonon environment that produces standard amplitude damping and heating processes.
Under the Born--Markov and secular approximations, the full dynamics of the system is described by the master equation~\cite{RevModPhys.73.357,PhysRevB.72.134519,Jaehne_2008}:
\begin{align}
\label{eq:masterArch}
\frac{d\rho}{dt}
&= -i[H(t), \rho]
+ \kappa \, \mathcal{D}[\sigma_-] \rho
+ \frac{\Gamma_\phi}{2} \, \mathcal{D}[\sigma_z] \rho
+ \gamma_- \, \mathcal{D}[a] \rho
+ \gamma_+ \, \mathcal{D}[a^\dagger] \rho.
\end{align}
Combining all coherent and dissipative contributions, the total Hamiltonian in the qubit eigenbasis takes the form
\begin{align}
H(t)
= \omega_m a^\dagger a
+ \frac{\omega_q}{2}\,\sigma_z
+ g_z\,\sigma_z (a + a^\dagger)
+ g_x\,\sigma_x (a + a^\dagger)
+ \Omega \cos(\omega_d t)\, \sigma_x .
\end{align}
In addition to the dissipative channels explicitly included in Eq.~\eqref{eq:EffME}, the electromagnetic environment also introduces a pure dephasing mechanism for the qubit, captured by the rate \( \Gamma_\phi \) as shown in Eq. (\ref{eq:masterArch}). Although this dephasing is not deliberately engineered, it arises naturally from the longitudinal coupling to environmental charge noise. However, it does not significantly affect the results presented in Sec.~\ref{sec:results}, since we operate in the regime \( \kappa \gg \{\gamma, \Gamma_\phi \}\), where the qubit rapidly relaxes to its ground state. In this limit, the additional dephasing has a negligible influence on the mechanical resonator dynamics. If dephasing is non-negligible, we need to include dephasing $\Gamma_\phi$  by replacing $\kappa/2 \rightarrow \kappa/2+\Gamma_\phi$ in the qubit response functions $S_\pm$ and $S_{2\pm}$; consequently, the qubit-induced rates in Eq.~(\ref{eq:EffME}) and the squeezing amplitude $\chi$ inherit this broadening.


\section{Derivation of the Effective Master Equation}\label{Appendix:B}
In this appendix, we present a detailed derivation of the second-order contributions to the effective master equation governing the reduced dynamics of the mechanical resonator, presented in Eq.~(\ref{eq:EffME}). 
Our goal is to adiabatically eliminate the qubit degrees of freedom and derive a closed master equation for the resonator, valid in the regime where the qubit rapidly relaxes to its steady state due to strong dissipation.
The effective interaction Hamiltonian, obtained via a unitary transformation [Eq.~(\ref{eq:SWHprime})], is transformed into the interaction picture with respect to the free evolution of the qubit and the resonator. In this frame, the interaction Hamiltonian decomposes as
\begin{equation}\label{eq:Hint2}
H_\text{int}(t) = H_1(t) + H_2(t) + H_3(t),
\end{equation}
where
\begin{align}
H_1(t) &= g_x ( \sigma_+ e^{i \omega_q t} + \sigma_- e^{-i \omega_q t} ) ( a e^{-i \omega_m t} + a^\dagger e^{i \omega_m t} ), \quad
H_2(t) = -g ( \sigma_- e^{-i \omega_q t} - \sigma_+ e^{i \omega_q t} ) ( a^{\dagger 2} e^{2 i \omega_m t} - a^2 e^{-2 i \omega_m t} ), \nonumber \\
H_3(t) &= \varepsilon ( e^{i \omega_d t} + e^{-i \omega_d t} ) ( \sigma_+ e^{i \omega_q t} + \sigma_- e^{-i \omega_q t} ).
\end{align}
To derive the effective resonator dynamics, we employ the Nakajima–Zwanzig projection operator formalism, as discussed in Sec.~\ref{sec:results}. The formalism yields
\begin{equation}
\frac{d}{dt} \mathcal{P} \rho(t) = \mathcal{P} \mathcal{L} \mathcal{P} \rho(t) + \int_0^\infty \! d\tau\, \mathcal{P} \mathcal{L}(t) e^{\mathcal{Q} \mathcal{L}_{0} \tau} \mathcal{Q} \mathcal{L}(t - s) \mathcal{P} \rho(t),
\end{equation}
where the projection operator is defined as \( \mathcal{P} \rho = \rho_q^{\mathrm{ss}} \otimes \rho_m(t) \), with \( \rho_q^{\mathrm{ss}} \) the steady-state density matrix of the qubit. In the strong qubit dissipation regime $\kappa \gg \gamma$, we can assume the qubit quickly relaxes to the ground state. Hence, in our calculation, we assume \( \rho_q^{\mathrm{ss}} = |g\rangle\langle g| \). In addition, \( \rho_m(t) \) is the reduced density matrix of the resonator, and 
the complementary projector is \( \mathcal{Q} = \mathbb{I} - \mathcal{P} \).
Since \( \operatorname{Tr}_q[H_\text{int}(t)\rho_q^{\mathrm{ss}}] = 0 \), the first-order term vanishes and the leading contribution comes from second-order processes. Expanding the interaction Liouvillian \( \mathcal{L}(t)\rho = -i [H_\text{int}(t), \rho] \), the second-order memory kernel becomes
\begin{equation}\label{eq:Mij}
M_{ij}(t) = - \operatorname{Tr}_q \left[ H_i(t), \int_0^\infty \! d\tau\, e^{\mathcal{L}_0 \tau} \left[ H_j(t - \tau), \rho_q^{\mathrm{ss}} \otimes \rho_m(t) \right] \right],
\end{equation}
where \( \mathcal{L}_0 \rho =  \kappa \mathcal{D}[\sigma_-] \rho \) governs the dissipative qubit dynamics, and \( \operatorname{Tr}_q \) denotes the partial trace over the qubit degrees of freedom.
Combining all second-order terms and intrinsic mechanical damping, the effective master equation for the resonator is given by
\begin{equation}
\frac{d}{dt} \rho_m(t) = \mathcal{L}_b \rho_m(t) + \sum_{i,j = 1}^{3} M_{ij}(t),
\end{equation}
where \( \mathcal{L}_b \) accounts for the intrinsic dissipation of the resonator due to its coupling to a thermal environment at temperature \( T \):
\begin{equation}
\mathcal{L}_b \rho_m = \gamma (n_\text{th} + 1) \mathcal{D}[a] \rho_m + \gamma n_\text{th} \mathcal{D}[a^\dagger] \rho_m.
\end{equation}
Among the nine second-order terms \( M_{ij} \), only three are non-zero \( M_{11} \), \( M_{22} \), and \( M_{23} \). Their derivations are provided in the following subsections.

\subsection{Single-phonon interaction: Derivation of $M_{11}$}

To derive $M_{11}$, first, we decompose the single-phonon interaction $H_1(t)$ in Eq. (\ref{eq:Hint2}) as
\begin{equation}
H_1(t) = g_x A_1(t) \otimes B_1(t), \qquad
A_1(t) = \sigma_+ e^{i \omega_q t} + \sigma_- e^{-i \omega_q t}, \qquad
B_1(t) = a e^{-i \omega_m t} + a^\dagger e^{i \omega_m t}.
\label{eq:H1t}
\end{equation}
Accordingly, the second-order Nakajima--Zwanzig kernel given in Eq. (\ref{eq:Mij}) reads
\begin{equation}
M_{11}(t)\rho_m
= 
-\operatorname{Tr}_q \!\left[
H_1(t), \int_{0}^{\infty}d\tau\ e^{\mathcal{L}_0 \tau}
\left[ H_1(t-\tau),\; |g\rangle\langle g| \otimes \rho_m \right]
\right],
\label{eq:M11_finite_time}
\end{equation}
with the (purely dissipative) reference Liouvillian $\mathcal{L}_0\rho = \kappa\,\mathcal{D}[\sigma_-]\rho$.
Acting on qubit operators, $e^{\mathcal{L}_0 \tau}\sigma_\pm = e^{-\kappa \tau/2}\sigma_\pm$.
Expanding the inner commutator of Eq. (\ref{eq:M11_finite_time}) using $\sigma_-|g\rangle=0$ and $|g\rangle\langle g|\,\sigma_+=0$ gives
\begin{equation}
Y(\tau) \equiv \left[ H_1(t-\tau),\; |g\rangle\langle g| \otimes \rho_m \right]
= g_x \Big(
\sigma_+ e^{i\omega_q(t-\tau)} \otimes B_1(t-\tau)\rho_m
- \sigma_- e^{-i\omega_q(t-\tau)} \otimes \rho_m B_1(t-\tau)
\Big).
\label{eq:inner_comm_expanded}
\end{equation}
Propagating with $e^{\mathcal{L}_0\tau}$ and forming the outer commutator with $H_1(t)$, tracing over the qubit, yields
\begin{align}
\operatorname{Tr}_q[H_1(t), e^{\mathcal{L}_0\tau}Y(\tau)]
= g_x^2 e^{-\kappa \tau/2} \Big\{&
e^{-i\omega_q\tau} \big( B_1(t) B_1(t-\tau)\rho_m - B_1(t-\tau)\rho_m B_1(t) \big)
\nonumber\\
&+ e^{+i\omega_q\tau} \big( - B_1(t) \rho_m B_1(t-\tau) + \rho_m B_1(t-\tau) B_1(t) \big)
\Big\}.
\label{eq:outer_comm_traced}
\end{align}
Using $B_1(t)= a e^{-i\omega_m t}+a^\dagger e^{i\omega_m t}$ and $B_1(t-\tau)= a e^{-i\omega_m (t-\tau)}+a^\dagger e^{i\omega_m (t-\tau)}$, the products $B_1(t)B_1(t-\tau)$ and other similar terms,  decompose into the operator monomials $a^2$, $aa^\dagger$, $a^\dagger a$, $a^{\dagger2}$ with phase factors $e^{\pm i\omega_m\tau}$ and global $e^{\pm 2i\omega_m t}$.
We define the response functions
\begin{equation}
S_\pm \equiv \int_0^\infty d\tau\, e^{-\kappa\tau/2}\, e^{-i(\omega_q\pm\omega_m)\tau}
= \frac{1}{\frac{\kappa}{2}+i\Delta_\pm}, \qquad \Delta_\pm = \omega_q\pm\omega_m.
\label{eq:Spm_Markov}
\end{equation}
Next, applying a rotating-wave approximation (RWA) at the mechanical frequency, and dropping the fast two-phonon sector $\propto e^{\pm 2i\omega_m t}$. The remaining number-like part of the kernel can be written as
\begin{align}
M_{11}\rho_m
= -g_x^2\Big[&
S_+\,(a a^\dagger\rho_m - a^\dagger\rho_m a)
+S_+^*( \rho_m a a^\dagger - a^\dagger\rho_m a)
+S_-\,(a^\dagger a\,\rho_m - a\,\rho_m a^\dagger)
+S_-^*(\rho_m a^\dagger a - a\,\rho_m a^\dagger)
\Big].
\label{eq:M11_preLindblad}
\end{align}
Grouping terms into canonical dissipators and a commutator gives the Lindblad form
\begin{equation}
M_{11}\rho_m
= - i\,\delta_1\,[a^\dagger a,\rho_m]+ \Gamma_1^-\,\mathcal{D}[a]\rho_m
+ \Gamma_1^+\,\mathcal{D}[a^\dagger]\rho_m,
\label{eq:M11_Lindblad}
\end{equation}
with the induced rates and Lamb shift is given by
\begin{equation}
\Gamma_1^\pm = 2 g_x^2\, \mathrm{Re}\,S_\pm,
\qquad
\delta_1 = g_x^2\, \big(\mathrm{Im}\,S_- + \mathrm{Im}\,S_+\big).
\label{eq:Rates_Shift}
\end{equation}
Eq. (\ref{eq:M11_Lindblad}) describes dissipative one-phonon processes induced by the fast-relaxing qubit, with rates determined by the spectral overlap of the drive and the qubit linewidth. The Lamb shift \(\delta_1\) arises from virtual excitation and de-excitation of the qubit and leads to a frequency renormalization of the mechanical mode. 

\subsection{Two-phonon interaction: Derivation of $M_{22}$}\label{subsec:M_22}
We now turn to the contribution from the two-phonon interaction Hamiltonian \(H_2(t)\), defined in Eq.~(\ref{eq:Hint2}). We express the interaction as
\begin{equation}
H_2(t) = g\,A_2(t)\otimes B_2(t),\qquad
A_2(t)=\sigma_+ e^{i\omega_q t}-\sigma_- e^{-i\omega_q t},\qquad
B_2(t)=a^{\dagger2} e^{2i\omega_m t}-a^2 e^{-2i\omega_m t}.
\label{eq:H2_def}
\end{equation}
The second-order NZ kernel given in Eq. (\ref{eq:Mij}) becomes
\begin{equation}
M_{22}(t)\rho_m
= -
\operatorname{Tr}_q\!\Big[
H_2(t),\int_{0}^{\infty} d\tau\, \ e^{\mathcal{L}_0\tau}\,[H_2(t-\tau),\,|g\rangle\!\langle g|\otimes\rho_m]
\Big].
\label{eq:M22_finite_time}
\end{equation}
Expanding the inner commutator using $\sigma_-|g\rangle=0$ and $|g\rangle\!\langle g|\,\sigma_+=0$ gives
\begin{equation}
[H_2(t-\tau),\,|g\rangle\!\langle g|\otimes\rho_m]
= g\Big(\sigma_+ e^{i\omega_q(t-\tau)}\otimes B_2(t-\tau)\rho_m
\;+\;\sigma_- e^{-i\omega_q(t-\tau)}\otimes \rho_m B_2(t-\tau)\Big).
\label{eq:M22_inner}
\end{equation}
Propagating Eq. (\ref{eq:M22_inner}) with $e^{\mathcal{L}_0\tau}$ and forming the outer commutator with $H_2(t)$ given in Eq. (\ref{eq:M22_finite_time}), then tracing over the qubit, we obtain
\begin{align}
M_{22}(t)\rho_m
= -\,g^2\!\int_0^{\infty}\! d\tau\, e^{-\kappa\tau/2}\Big[
e^{-i\omega_q\tau}\,[B_2(t-\tau)\rho_m,\ B_2(t)]
+ e^{+i\omega_q\tau}\,[B_2(t),\ \rho_m B_2(t-\tau)]
\Big].
\label{eq:M22_comm_blocks}
\end{align}
With $B_2(t)=a^{\dagger2}e^{2i\omega_m t}-a^2e^{-2i\omega_m t}$ and
$B_2(t-\tau)=a^{\dagger2}e^{2i\omega_m t}e^{-2i\omega_m\tau}-a^2e^{-2i\omega_m t}e^{+2i\omega_m\tau}$, all products decompose into quartic monomials
$a^{\dagger2}a^{\dagger2}$, $a^{\dagger2}a^2$, $a^2a^{\dagger2}$, $a^2a^2$ with phases $e^{\pm 2i\omega_m\tau}$ and global factors $e^{\pm 4i\omega_m t}$.
We define the two-phonon response functions
\begin{equation}
S_{2\pm}
=\int_0^{\infty}\! d\tau\, e^{-\kappa\tau/2} e^{-i(\omega_q\pm 2\omega_m)\tau}
=\frac{1}{\frac{\kappa}{2}+i\Delta_{2\pm}},
\qquad
\Delta_{2\pm}:=\omega_q\pm 2\omega_m.
\label{eq:S2pm}
\end{equation}
Collecting the non-oscillatory terms only in Eq. (\ref{eq:M22_comm_blocks}) and reorganizing into canonical dissipators plus a Hamiltonian contribution yields
\begin{equation}
M_{22}(t)\rho_m
=- i\,[H_{22},\rho_m] + \Gamma_2^-\,\mathcal{D}[a^2]\rho_m
+\Gamma_2^+\,\mathcal{D}[a^{\dagger2}]\rho_m,
\label{eq:M22_stationary}
\end{equation}
with dissipative rates and coherent (Lamb/Kerr) Hamiltonian
\begin{equation}
\Gamma_2^\pm = 2\,g^2\,\mathrm{Re}\,S_{2\pm}, \,\,\text{and}\quad H_{22}
= g^2\Big(\mathrm{Im}\,S_{2-}\,a^{\dagger2}a^2+\mathrm{Im}\,S_{2+}\,a^{2}a^{\dagger2}\Big)
= \delta_k\,\hat n^2+\delta_2\,\hat n+\mathrm{const},
\label{eq:Gamma2}
\end{equation}
where $\hat n=a^\dagger a$ and we used $a^{\dagger2}a^2=\hat n^2-\hat n$, $a^{2}a^{\dagger2}=\hat n^2+3\hat n+2$.
The coefficients are
\begin{equation}
\delta_k=g^2\big(\mathrm{Im}\,S_{2-}+\mathrm{Im}\,S_{2+}\big),\qquad
\delta_2=g^2\big(-\mathrm{Im}\,S_{2-}+3\,\mathrm{Im}\,S_{2+}\big),
\label{eq:dk_d2}
\end{equation}
with the constant in \eqref{eq:Gamma2} physically irrelevant.

\subsection{Two-phonon–drive cross term: Derivation of \texorpdfstring{$M_{23}$}{M23}}

We now evaluate the cross kernel that couples the two-phonon interaction \(H_2(t)\) to the external drive \(H_3(t)\). The relevant interaction-picture operators (cf.\ Eq.~(\ref{eq:Hint2})) are
\begin{align}
H_2(t)&=g\,A_2(t)\otimes B_2(t), \qquad
A_2(t) =\sigma_+e^{i\omega_q t}-\sigma_-e^{-i\omega_q t}, \qquad
B_2(t) =a^{\dagger2}e^{2i\omega_m t}-a^2 e^{-2i\omega_m t}, \nonumber \\ 
H_3(t)&=\varepsilon\,(e^{i\omega_d t}+e^{-i\omega_d t})\big(\sigma_+e^{i\omega_q t}+\sigma_-e^{-i\omega_q t}\big),
\label{Eq:H_2(t)}
\end{align}
Eq. (\ref{eq:Mij}) reads
\begin{equation}
M_{23}(t)\rho_m
=
-\operatorname{Tr}_q\!\left[
H_2(t),\;
\int_0^\infty\! d\tau\;
e^{\mathcal{L}_0\tau}\,
\big[H_3(t-\tau),\,|g\rangle\langle g|\otimes\rho_m\big]
\right].
\label{eq:M23_def}
\end{equation}
Writing out the drive term,
\begin{align}
H_3(t-\tau)
&=\varepsilon\Big\{
\sigma_+\!\Big(e^{i(\omega_d+\omega_q)(t-\tau)}+e^{i(\omega_q-\omega_d)(t-\tau)}\Big)
+\sigma_-\!\Big(e^{i(\omega_d-\omega_q)(t-\tau)}+e^{-i(\omega_d+\omega_q)(t-\tau)}\Big)
\Big\},
\end{align}
\begin{table}[h!]
\centering
\caption{Parameters entering the effective master equation (\ref{eq:EffME}) and their physical roles.}
\begin{tabular}{lll}
\hline\hline
\textbf{Symbol} & \textbf{Expression} & \textbf{Physical Role} \\ \hline
$\Gamma_1^{-}$ & $2g_x^{2}\,\text{Re}\,S_{-}$ & Qubit‑mediated one‑phonon absorption rate (cooling) \\
$\Gamma_1^{+}$ & $2g_x^{2}\,\text{Re}\,S_{+}$ & Qubit‑mediated one‑phonon emission rate (heating) \\
$\Gamma_2^{-}$ & $2g^{2}\,\text{Re}\,S_{2-}$ & Qubit‑mediated two‑phonon absorption rate\\
$\Gamma_2^{+}$ & $2g^{2}\,\text{Re}\,S_{2+}$ & Qubit‑mediated two‑phonon emission rate \\
$\gamma$       & ― (intrinsic) & Bare mechanical damping rate set by the thermal bath \\
$\kappa$       & ― (intrinsic) & Qubit decay rate \\
$n_{\mathrm{th}}$ & ― (intrinsic) & Mean thermal phonon number of the resonator’s environment \\
$\delta_{1}$   & $g_x^{2}\!\bigl(\text{Im}\,S_{-}+\text{Im}\,S_{+}\bigr)$ 
               & One‑phonon Lamb shift of the resonator frequency \\
$\delta_{2}$   & $g^{2}\!\bigl(-\text{Im}\,S_{2-}+3\,\text{Im}\,S_{2+}\bigr)$ 
               & Two‑phonon Lamb (frequency‑renormalization) shift \\
$\delta_{k}$   & $g^{2}\!\bigl(\text{Im}\,S_{2-}+\text{Im}\,S_{2+}\bigr)$ 
               & Kerr‑type nonlinearity ($\propto\hat n^{2}$) \\
$\chi$         & $\displaystyle-\frac{i \varepsilon g}{\kappa/2 - i(\omega_q - \omega_d)}.$ 
               & Coherent two‑phonon squeezing amplitude generated by the drive \\
\hline\hline
\end{tabular}
\label{tab:EffME_params}
\end{table}
and using \(\sigma_-|g\rangle=0\) and \(|g\rangle\langle g|\sigma_+=0\), the inner commutator of Eq. (\ref{eq:M23_def}) is given as
\begin{equation}
[H_3(t-\tau),\,|g\rangle\langle g|\otimes\rho_m]
=\varepsilon\Big(\sigma_+\,F_+(t,\tau)-\sigma_-\,F_-(t,\tau)\Big)\otimes\rho_m,
\label{eq:inner}
\end{equation}
where
\begin{align}
F_\pm(t,\tau) = e^{\pm i(\omega_q+\omega_d)(t-\tau)} + e^{\pm i(\omega_q-\omega_d)(t-\tau)}.
\label{eq:F_pm}
\end{align}
Acting with \(e^{\mathcal{L}_0\tau}\) on Eq. (\ref{eq:inner}) produces
\begin{equation}
Z (\tau) \equiv e^{\mathcal{L}_0\tau}[H_3(t-\tau),\,|g\rangle\langle g|\otimes\rho_m]
=\varepsilon\,e^{-\kappa\tau/2}\Big(\sigma_+\,F_+(t,\tau)-\sigma_-\,F_-(t,\tau)\Big)\otimes\rho_m.
\label{eq:Z}
\end{equation}
We now compute the outer commutator \([H_2(t),\,Z(\tau)]\) in Eq. (\ref{eq:M23_def})  and trace out the qubit. Using
\begin{align}
A_2(t)\sigma_+&=-\,|g\rangle\langle g|\,e^{-i\omega_q t}, \quad
\sigma_+A_2(t) =-\,|e\rangle\langle e|\,e^{-i\omega_q t}, \quad
A_2(t)\sigma_- =+\,|e\rangle\langle e|\,e^{+i\omega_q t}, \quad
\sigma_-A_2(t) =+\,|g\rangle\langle g|\,e^{+i\omega_q t},
\end{align}
a straightforward expansion yields
\begin{equation}
\operatorname{Tr}_q\!\big[H_2(t),\,Z(\tau)\big]
=-\,g\,\varepsilon\,e^{-\kappa\tau/2}\,\mathcal{G}(t,\tau)\,[B_2(t),\rho_m],
\label{eq:traceq}
\end{equation}
where the scalar prefactor is
\begin{equation}
\mathcal{G}(t,\tau)
= e^{-i\omega_q t}F_+(t,\tau)+e^{+i\omega_q t}F_-(t,\tau).
\end{equation}
Using Eq. (\ref{eq:traceq}) in Eq. (\ref{eq:M23_def}) we obtain
\begin{equation}
M_{23}(t)\rho_m
= - g\,\varepsilon\int_0^\infty\! d\tau\, e^{-\kappa\tau/2}\,\mathcal{G}(t,\tau)\,[\rho_m,B_2(t)].
\label{eq:M23_comm}
\end{equation}
Introducing the response integral
\begin{equation}
S(\Delta)=\int_0^\infty d\tau\,e^{-\kappa\tau/2}e^{-i\Delta\tau}
=\frac{1}{\kappa/2+i\Delta},
\end{equation}
Eq.~(\ref{eq:M23_comm}) becomes
\begin{align}
M_{23}(t)\rho_m
&= g\,\varepsilon\Big\{
e^{+i\omega_d t}\big[S(\omega_d+\omega_q)+S(\omega_d-\omega_q)\big]
+e^{-i\omega_d t}\big[S^*(\omega_d-\omega_q)+S^*(\omega_d+\omega_q)\big]
\Big\}\,[\rho_m,B_2(t)].
\label{eq:M23_preSplit}
\end{align}
Using \(B_2(t)=a^{\dagger2}e^{2i\omega_m t}-a^2 e^{-2i\omega_m t}\), we obtain
\begin{align}
M_{23}(t)\rho_m
&=\Xi_+(t)\,[\rho_m,a^{\dagger2}] + \Xi_-(t)\,[\rho_m,a^2],
\end{align}
with
\begin{align}
\Xi_+(t)
&= -\,g\,\varepsilon\,e^{2i\omega_m t}\Big\{
e^{+i\omega_d t}\big[S(\omega_d+\omega_q)+S(\omega_d-\omega_q)\big]
+e^{-i\omega_d t}\big[S^*(\omega_d-\omega_q)+S^*(\omega_d+\omega_q)\big]
\Big\},\\
\Xi_-(t)
&= \,g\,\varepsilon\,e^{-2i\omega_m t}\Big\{
e^{+i\omega_d t}\big[S(\omega_d+\omega_q)+S(\omega_d-\omega_q)\big]
+e^{-i\omega_d t}\big[S^*(\omega_d-\omega_q)+S^*(\omega_d+\omega_q)\big]
\Big\}.
\end{align}
This expression is exact: it contains no rotating-wave approximation and exhibits a purely commutator structure, i.e., it generates coherent two-phonon processes and no dissipators.
When the drive is set to the parametric resonance \(\omega_d=2\omega_m\), the global factors \(e^{\pm i\omega_d t}e^{\mp 2i\omega_m t}\) cancel, and the coefficients become time independent. Furthermore, for a near-resonant qubit, \(|\omega_q-\omega_d|\ll\omega_q+\omega_d\), the term with \(S(\omega_d-\omega_q)=S(-(\omega_q-\omega_d))=S^*(\omega_q-\omega_d)\) dominates over the far-detuned \(S(\omega_d+\omega_q)\). Retaining the dominant contribution yields a purely Hamiltonian form
\begin{equation}
M_{23}\rho_m=-i[H_{23},\rho_m],\qquad
H_{23}=\chi\,a^2+\chi^*\,a^{\dagger2},
\end{equation}
with the squeezing amplitude
\begin{equation}
\chi
=
-\,i\,\frac{g\,\varepsilon}{\kappa/2 - i\Delta_d}, \qquad
\Delta_d=\omega_q-\omega_d.
\end{equation}
Thus, the drive--nonlinear cross term contributes a coherent two-phonon squeezing interaction that, together with the dissipative two-phonon channel from \(M_{22}\), enables stabilization of an even (odd)-parity steady state.  Parity is preserved provided $\Gamma_{2}^{-} \gg (\Gamma_{1}^{-}, \Gamma_{1}^{+}, \gamma_{-}, \gamma_{+})$. In practice, this is ensured by operating near the two-phonon resonance,
\(|\Delta_{2-}| \ll |\Delta_{2+}|\),
and choosing a sufficiently large qubit linewidth \(\kappa\) so that the
single-phonon rates \(\Gamma_{1}^{\pm}\) are suppressed relative to \(\Gamma_{2}^{-}\).

{\color{black}
\subsection{Why Only Three Second-Order Kernels Survive}

To make the selection of second–order contributions fully transparent, this subsection provides explicit calculations showing which Nakajima--Zwanzig kernels in Eq.~\eqref{eq:Mij} vanish and why. Starting from the nine kernels \(M_{ij}\) generated by \(H_1\), \(H_2\), and \(H_3\), and building on the explicit evaluations of \(M_{11}\), \(M_{22}\), and \(M_{23}\) in the preceding subsections, we demonstrate that only these three kernels survive. The remaining six fall into three natural groups, each eliminated by the same mechanism. Group~I (outer drive commutator), \(M_{31}\), \(M_{32}\), and \(M_{33}\), vanishes because the partial trace of a qubit--only commutator is identically zero. Group~II (mixed one-- and two--phonon vertices), \(M_{12}\) and \(M_{21}\), reduces to cubic mechanical monomials that carry only fast phases at \(\pm\omega_m\) or \(\pm 3\omega_m\) and therefore average out under the RWA. Group~III (one--phonon and drive), \(M_{13}\), is off--resonant; with \(\omega_d \simeq 2\omega_m\), its residual detunings are \(\omega_d \pm \omega_m\), so no near--stationary contribution remains after coarse--graining. Throughout, the eliminations follow two explicit criteria: (i) a rotating--wave rule, whereby only near--zero net frequencies are retained, and (ii) a mechanical selection rule, whereby odd mechanical monomials do not yield stationary generators in the coarse--grained dynamics.

\subsubsection{Group I: Vanishing kernels with an outer drive commutator (\texorpdfstring{$M_{31}$, $M_{32}$, $M_{33}$}{M31, M32, M33})}

Here we show that all kernels whose outer commutator involves the drive Hamiltonian $H_3(t)$ vanish:
\begin{equation}
M_{31}(t)\rho_m=0,\qquad M_{32}(t)\rho_m=0,\qquad M_{33}(t)\rho_m=0.
\end{equation}
The cancellations originate from the fact that, after expanding the inner commutator in Eq. \eqref{eq:Mij} and propagating with $e^{\mathcal L_0\tau}$, the outer commutator with $H_3(t)$ produces a factor of $\sigma_z$ acting only on the qubit, whose partial trace is zero. 
In what follows, we repeatedly use the cyclicity identity under a partial trace: for any qubit operator $A_q$ and any bipartite $X=\sum_k O_{q,k}\otimes O_{m,k}$,
\begin{equation}
\operatorname{Tr}_q\big([A_q,X]\big)=0,
\label{eq:G1_cyclicity}
\end{equation}
and the elementary property $\operatorname{Tr}_q(\sigma_z)=0$.

(i) {\it{Kernel $M_{31}$}}: $H_3$ outside, $H_1$ inside.
To evaluate \(M_{31}(t)\), we start from Eq.~\eqref{eq:Mij} with \(i=3\) and \(j=1\), and take the qubit steady state as \(\rho_q^{\mathrm{ss}} = |g\rangle\langle g|\). This gives
\begin{equation}
M_{31}(t)\rho_m
=-
\operatorname{Tr}_q\!\left[
H_3(t),\int_0^\infty d\tau\ e^{\mathcal L_0\tau}\,
\big[H_1(t-\tau),\,|g\rangle\langle g|\otimes\rho_m\big]
\right].
\label{eq:G1_M31_start}
\end{equation}
For the inner commutator, we substitute \(H_1(t)\) from Eq. \eqref{eq:H1t}, this yields
\begin{equation}
C_{31}(\tau)=\big[H_1(t-\tau),\,|g\rangle\langle g|\otimes\rho_m\big]
=g_x\Big(\sigma_+e^{i\omega_q (t-\tau)}\otimes B_1(t-\tau)\rho_m
-\sigma_-e^{-i\omega_q (t-\tau)}\otimes \rho_m B_1(t-\tau)\Big).
\label{eq:G1_M31_inner}
\end{equation}
Next, we propagate \(C_{31}(\tau)\) under the Liouvillian \(\mathcal L_0\), this gives
\begin{equation}
e^{\mathcal L_0\tau}C_{31}(\tau)
=g_x e^{-\kappa\tau/2}\Big(\sigma_+e^{i\omega_q (t-\tau)}\otimes B_1(t-\tau)\rho_m
-\sigma_-e^{-i\omega_q (t-\tau)}\otimes \rho_m B_1(t-\tau)\Big).
\label{eq:G1_M31_prop}
\end{equation}
We now evaluate the outer commutator in Eq. \eqref{eq:G1_M31_start} with the drive Hamiltonian \(H_3(t)\) given in Eq. \eqref{eq:Hint2}. Using
\begin{equation}
[H_3(t),\sigma_+]=-\,\varepsilon E_d(t)\,e^{-i\omega_q t}\sigma_z,\qquad
[H_3(t),\sigma_-]=\,\varepsilon E_d(t)\,e^{+i\omega_q t}\sigma_z,
\label{eq:G1_H3_comm}
\end{equation}
we obtain
\begin{equation}
\big[H_3(t),e^{\mathcal L_0\tau}C_{31}(\tau)\big]
=-\,g_x\varepsilon\,E_d(t)\,e^{-\kappa\tau/2}
\Big(e^{-i\omega_q\tau}\,\sigma_z\otimes B_1(t-\tau)\rho_m
+e^{+i\omega_q\tau}\,\sigma_z\otimes \rho_m B_1(t-\tau)\Big),
\label{eq:G1_M31_outer}
\end{equation}
here \(E_d(t) = e^{i\omega_d t} + e^{-i\omega_d t}\). Taking the partial trace over the qubit in Eq. \eqref{eq:G1_M31_outer} and using
\begin{equation}
    \operatorname{Tr}_q(\sigma_z\otimes X_m)=\operatorname{Tr}_q(\sigma_z)\, X_m=0,
    \label{eq:ptrace0}
\end{equation}
we immediately obtain
\begin{equation}
\,M_{31}(t)\rho_m=0.\,
\label{eq:G1_M31_zero}
\end{equation}

(ii) \emph{Kernel $M_{32}$}: $H_3$ outside, $H_2$ inside.
Analogously,
\begin{equation}
M_{32}(t)\rho_m
=-
\operatorname{Tr}_q\!\left[
H_3(t),\int_0^\infty d\tau\ e^{\mathcal L_0\tau}\,
\big[H_2(t-\tau),\,|g\rangle\langle g|\otimes\rho_m\big]
\right],
\label{eq:G1_M32_start}
\end{equation}
with $H_2(t)$ given in Eq. \eqref{Eq:H_2(t)}. Following the similar procedure as in $M_{31}$, the inner commutator becomes
\begin{equation}
C_{32}(\tau)=\big[H_2(t-\tau),\,|g\rangle\langle g|\otimes\rho_m\big]
=g\Big(\sigma_+e^{i\omega_q (t-\tau)}\otimes B_2(t-\tau)\rho_m
+\sigma_-e^{-i\omega_q (t-\tau)}\otimes \rho_m B_2(t-\tau)\Big).
\label{eq:G1_M32_inner}
\end{equation}
Applying $e^{\mathcal L_0\tau}$ and Eq.~\eqref{eq:G1_H3_comm}, the outer commutator in Eq. \eqref{eq:G1_M32_start} modifies to
\begin{equation}
\big[H_3(t),e^{\mathcal L_0\tau}C_{32}(\tau)\big]
=-\,g\,\varepsilon\,E_d(t)\,e^{-\kappa\tau/2}
\Big(e^{-i\omega_q\tau}\,\sigma_z\otimes B_2(t-\tau)\rho_m
- e^{+i\omega_q\tau}\,\sigma_z\otimes \rho_m B_2(t-\tau)\Big).
\label{eq:G1_M32_outer}
\end{equation}
Taking the partial trace over the qubit and using the property given in Eq. \eqref{eq:ptrace0}, we obtain
\begin{equation}
M_{32}(t)\rho_m=0.\,
\label{eq:G1_M32_zero}
\end{equation}

(iii) \emph{Kernel $M_{33}$}: $H_3$ outside, $H_3$ inside.
Finally, we consider the kernel with the drive Hamiltonian acting on both sides. Starting from
\begin{equation}
M_{33}(t)\rho_m
=-
\operatorname{Tr}_q\!\left[
H_3(t),\int_0^\infty d\tau\ e^{\mathcal L_0\tau}\,
\big[H_3(t-\tau),\,|g\rangle\langle g|\otimes\rho_m\big]
\right],
\label{eq:G1_M33_start}
\end{equation}
and using $H_3(t)$ given in Eq. \eqref{eq:Hint2}, the inner commutator followed by propagation under $e^{\mathcal L_0\tau}$ gives
\begin{equation}
e^{\mathcal L_0\tau}\big[H_3(t-\tau),\,|g\rangle\langle g|\otimes\rho_m\big]
=\varepsilon\,e^{-\kappa\tau/2}\Big(\sigma_+F_+(t,\tau)-\sigma_-F_-(t,\tau)\Big)\otimes\rho_m,
\label{eq:G1_M33_prop}
\end{equation}
with $F_\pm(t,\tau)$ is given in Eq. \eqref{eq:F_pm}. Taking the outer commutator in Eq. \eqref{eq:G1_M33_start} and using Eq.~\eqref{eq:G1_H3_comm}, we obtain
\begin{equation}
\big[H_3(t),\,e^{\mathcal L_0\tau}[H_3(t-\tau),\,|g\rangle\langle g|\otimes\rho_m]\big]
=-\,2\,\varepsilon^2\,E_d(t)\,E_d(t-\tau)\,e^{-\kappa\tau/2}\cos(\omega_q\tau)\,
\sigma_z\otimes\rho_m.
\label{eq:G1_M33_outer}
\end{equation}
Finally, taking the partial trace over the qubit and invoking  Eq. \eqref{eq:ptrace0}, we arrive at
\begin{equation}
M_{33}(t)\rho_m=0.
\label{eq:G1_M33_zero}
\end{equation}

\subsubsection{Group II: Mixed one-- and two--phonon vertices (\texorpdfstring{$M_{12}$, $M_{21}$}{M12, M21})}

The mixed kernels \(M_{12}\) and \(M_{21}\), which couple the single-phonon and two-phonon vertices, vanish after RWA. Physically, every qubit contraction produces cubic mechanical monomials \(\{a^{\dagger 2}a,\;a^{2}a^\dagger,\;a^{3},\;a^{\dagger 3}\}\) that carry net oscillation phases at \(\pm\omega_m\) or \(\pm 3\omega_m\). These fast phases remove any time-independent contribution, so no stationary Lindblad or coherent term survives in the RWA. Below we show this explicitly for both orderings, using Eqs.~\eqref{eq:Mij}, \eqref{eq:H1t}, and \eqref{eq:H2_def}.

\paragraph*{Kernel \(M_{12}\): outer \(H_1\), inner \(H_2\).}
We start from Eq.~\eqref{eq:Mij} with \(i=1\) and \(j=2\), which gives
\begin{equation}
M_{12}(t)\rho_m
=-
\operatorname{Tr}_q\!\left[
H_1(t),\;
\int_0^\infty d\tau\, e^{\mathcal L_0\tau}\,
\big[H_2(t-\tau),\,|g\rangle\langle g|\otimes\rho_m\big]
\right].
\label{eq:G2_M12_start}
\end{equation}
Using \(H_2(t)\) given in Eq.~\eqref{eq:H2_def}, the inner commutator becomes
\begin{equation}
\big[H_2(t-\tau),\,|g\rangle\langle g|\otimes\rho_m\big]
=
g\left(
\sigma_+ e^{i\omega_q(t-\tau)}\!\otimes\! B_2(t-\tau)\rho_m
+\sigma_- e^{-i\omega_q(t-\tau)}\!\otimes\! \rho_m B_2(t-\tau)
\right).
\label{eq:G2_M12_inner}
\end{equation}
Propagating the qubit operators under $e^{\mathcal L_0\tau}$ and then forming the outer commutator with \(H_1(t)\) yields, after tracing out the qubit,
\begin{equation}
M_{12}(t)\rho_m
=
g\,g_x
\!\int_0^\infty\! d\tau\, e^{-\kappa\tau/2}
\left(
e^{-i\omega_q\tau}\,[B_1(t),\,B_2(t-\tau)\rho_m]
+
e^{+i\omega_q\tau}\,[B_1(t),\,\rho_m B_2(t-\tau)]
\right).
\label{eq:G2_M12_reduced}
\end{equation}
Every product in Eq.~\eqref{eq:G2_M12_reduced} is cubic in the mechanical operators and carries global phases \(e^{\pm i\omega_m t}\) or \(e^{\pm 3 i\omega_m t}\), together with \(\tau\)–dependent factors \(e^{-i(\omega_q\pm 2\omega_m)\tau}\). The latter define the two--phonon response functions \(S_{2\pm}\) in Eq.~\eqref{eq:S2pm}. Collecting terms yields a linear combination of
\([a^{\dagger 2}a,\rho_m]\), \([a^{2}a^\dagger,\rho_m]\), \([a^3,\rho_m]\), and \([a^{\dagger 3},\rho_m]\),
each multiplied by \(e^{\pm i\omega_m t}\) or \(e^{\pm 3i\omega_m t}\).
By the secular test, these rapidly oscillating contributions average to zero under coarse--graining, so no time–independent generator remains:
\begin{equation}
\,M_{12}(t)\rho_m=0. 
\label{eq:G2_M12_zero}
\end{equation}

\paragraph*{Kernel \(M_{21}\): outer \(H_2\), inner \(H_1\).}
Interchanging the roles of \(H_1\) and \(H_2\) we start from
\begin{equation}
M_{21}(t)\rho_m
=-
\operatorname{Tr}_q\!\left[
H_2(t),\;
\int_0^\infty d\tau\, e^{\mathcal L_0\tau}\,
\big[H_1(t-\tau),\,|g\rangle\langle g|\otimes\rho_m\big]
\right].
\label{eq:G2_M21_start}
\end{equation}
The inner commutator now reads
\begin{equation}
\big[H_1(t-\tau),\,|g\rangle\langle g|\otimes\rho_m\big]
=
g_x\Big(
\sigma_+ e^{i\omega_q(t-\tau)}\!\otimes\! B_1(t-\tau)\rho_m
-\sigma_- e^{-i\omega_q(t-\tau)}\!\otimes\! \rho_m B_1(t-\tau)
\Big),
\label{eq:G2_M21_inner}
\end{equation}
now propagating under \(e^{\mathcal{L}_0 \tau}\), evaluating the outer commutator with \(H_2(t)\) defined in Eq.~\eqref{Eq:H_2(t)}, and tracing over the qubit finally yield\begin{equation}
M_{21}(t)\rho_m
=
g\,g_x
\!\int_0^\infty\! d\tau\, e^{-\kappa\tau/2}
\left(
e^{-i\omega_q\tau}\,[B_2(t),\,B_1(t-\tau)\rho_m]
+
e^{+i\omega_q\tau}\,[B_2(t),\,\rho_m B_1(t-\tau)]
\right).
\label{eq:G2_M21_reduced}
\end{equation}
Here \(B_1(t-\tau)\) contributes \(\tau\)–phases \(e^{\mp i\omega_m\tau}\), so the response functions are \(S_\pm\) of Eq.~\eqref{eq:Spm_Markov}. As in the previous case, every mechanical product is cubic and carries global factors \(e^{\pm i\omega_m t}\) or \(e^{\pm 3i\omega_m t}\).
By the same stationarity criterion, the coarse--grained kernel vanishes:
\begin{equation}
M_{21}(t)\rho_m=0.
\label{eq:G2_M21_zero}
\end{equation}
In summary, both mixed one– and two–phonon kernels \(M_{12}\) and \(M_{21}\) reduce exclusively to fast, cubic mechanical sectors and therefore contribute no stationary term to the effective generator.

\subsubsection{Group III: One--phonon vertice and drive (\texorpdfstring{$M_{13}$}{M13})}

\begin{figure}[t]
   \centering
   \includegraphics[width=\textwidth,
                    height=0.36\paperheight,
                    keepaspectratio,
                    trim=6pt 2pt 6pt 2pt,clip]{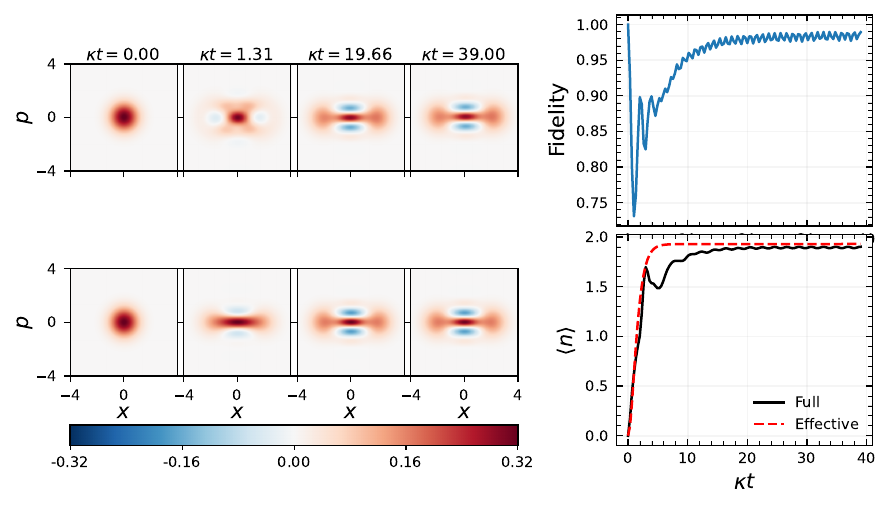}
   \caption{\small 
   Left: Wigner snapshots of the mechanical state from the full driven qubit–mechanics model (top row) and from the effective master equation (bottom row) at $\kappa t=0,\,1.31,\,19.66,\,39.00$. Both  descriptions exhibit the bifurcation into two lobes and the development of parity-protected interference fringes, converging to a steady cat state.
   Right: (top) Uhlmann fidelity $F(\rho_{\mathrm{full}}(t),\rho_{\mathrm{eff}}(t))$ between the full model and the reduced effective model; (bottom) mean phonon number $\langle n\rangle(t)$ from the full (solid black) and effective (red dashed) dynamics.
   The effective description closely tracks the full evolution and attains $F\gtrsim 0.98$ at late times.
   Parameters: $g_z/2\pi=6~\mathrm{MHz}$, $g_x=0.1\,g_z$, $\omega_m/2\pi=100~\mathrm{MHz}$, $\kappa/2\pi=100~\mathrm{kHz}$, $\gamma/2\pi=15~\mathrm{Hz}$, $\varepsilon = 2g$, $\omega_q=\omega_d=2\omega_m$, and $n_{\mathrm{th}}=0$.}
   \label{fig:fig2}
 \end{figure}

The last mixed kernel, \(M_{13}\), couples the single--phonon vertex to the classical drive. After RWA this kernel does not contribute to the effective generator. In contrast to Group~I, the cancellation here does not follow from a trace--of--commutator identity; instead, it is caused by off--resonant oscillations at frequencies \(\omega_d \pm \omega_m\). In what follows we make this explicit, starting from the general second--order kernel \eqref{eq:Mij} together with Eqs.~\eqref{eq:Hint2} and \eqref{eq:H1t}.
Setting $i=1$, and $j=3$ in Eq.~\eqref{eq:Mij} we obtain
\begin{equation}
M_{13}(t)\rho_m
=-
\operatorname{Tr}_q\!\left[
H_1(t),\;
\int_0^\infty d\tau\, e^{\mathcal L_0\tau}\,
\big[H_3(t-\tau),\,|g\rangle\langle g|\otimes\rho_m\big]
\right].
\label{eq:G3_M13_start}
\end{equation}
Using $H_3 (t)$ definition given in Eq \eqref{eq:Hint2}, the inner commutator of Eq. \eqref{eq:G3_M13_start} reads
\begin{equation}
\big[H_3(t-\tau),\,|g\rangle\langle g|\otimes\rho_m\big]
=\varepsilon\Big(\sigma_+F_+(t,\tau)-\sigma_-F_-(t,\tau)\Big)\otimes\rho_m,
\label{eq:G3_inner}
\end{equation}
here, $F_\pm(t,\tau)$ is given in Eq. \eqref{eq:F_pm}. Propagating the inner commutator in Eq. \eqref{eq:G3_inner} with $e^{\mathcal L_0\tau}$ and forming the outer commutator with $H_1(t)$ [Eq.~\eqref{eq:H1t}], the qubit trace gives
\begin{equation}
M_{13}(t)\rho_m
=
-g_x\varepsilon\!\int_0^\infty\! d\tau\, e^{-\kappa\tau/2}\,
\Big(e^{-i\omega_q t}F_+(t,\tau)-e^{+i\omega_q t}F_-(t,\tau)\Big)\,[B_1(t),\rho_m].
\label{eq:G3_commB1}
\end{equation}
We note that the \(M_{13}\) can only generate Hamiltonian--like contributions and cannot produce dissipators. Using the standard response integral in  Eq.~\eqref{eq:Spm_Markov}, we obtain the closed expression of Eq. \eqref{eq:G3_commB1}
\begin{equation}
M_{13}(t)\rho_m
=
-g_x\varepsilon\,\Big\{
e^{+i\omega_d t}\big[S(\omega_q+\omega_d)-S^*(\omega_q-\omega_d)\big]
+e^{-i\omega_d t}\big[S(\omega_q-\omega_d)-S^*(\omega_q+\omega_d)\big]
\Big\}\,[B_1(t),\rho_m].
\label{eq:G3_M13_exact}
\end{equation}
Substituting $B_1(t)$ into Eq. \eqref{eq:G3_M13_exact} shows that every contribution to \(M_{13}(t)\rho_m\) carries net phases
$e^{\pm i(\omega_d\pm\omega_m)t}$.
At the operating point $\omega_d\simeq 2\omega_m$, these combinations reduce to
$\{\omega_m,3\omega_m\}$, i.e.\ fast oscillations with no component near zero frequency.
Under the RWA applied throughout this appendix, such rapidly rotating factors average to zero. Therefore the coarse--grained kernel vanishes:
\begin{equation}
M_{13}(t)\rho_m=0.
\label{eq:G3_M13_zero}
\end{equation}
}

\setlength\textfloatsep{10pt plus 2pt minus 2pt}
\setlength\intextsep{10pt plus 2pt minus 2pt}
\setlength\abovecaptionskip{4pt}
\setlength\belowcaptionskip{2pt}

\begin{figure*}[t]
    \centering
    \includegraphics[width=0.85\columnwidth]{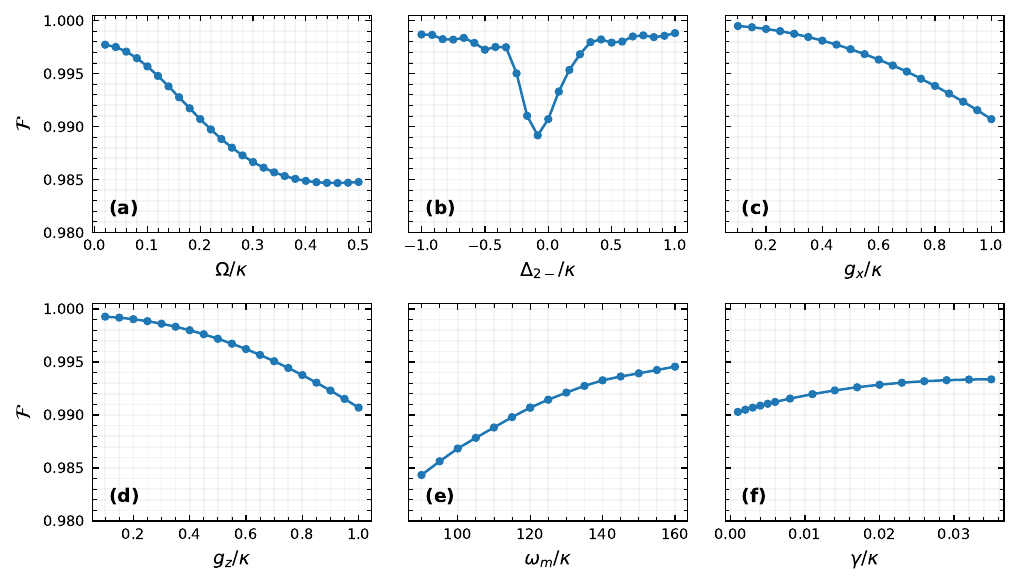}
    \caption{
       Fidelity $\mathcal{F}$ between the reduced oscillator state obtained from the full driven qubit--oscillator master equation~\eqref{eq:master} and the oscillator state obtained from the effective oscillator-only master equation~\eqref{eq:EffME}, evaluated at a fixed comparison time $\kappa t^\ast = 20$. Each panel shows $\mathcal{F}$ as a function of a single control parameter: (a) drive amplitude $\Omega$, (b) two-phonon detuning $\Delta_{2-} = \omega_q - 2\omega_m$, (c) transverse coupling $g_x$, (d) longitudinal coupling $g_z$, (e) mechanical frequency $\omega_m$, and (f) mechanical damping rate $\gamma$. The vertical axis is plotted on a linear scale and zoomed near $\mathcal{F}=1$ to highlight small deviations between the two descriptions. In each panel, only the indicated parameter is varied while all others are held fixed. (a) $(\Delta_{2-}, g_x, g_z, \omega_m, \gamma) = (0,\ 1.0,\ 1.0,\ 120,\ 3\times 10^{-3})$, (b) $(\Omega, g_x, g_z, \omega_m, \gamma) = (0.20,\ 1.0,\ 1.0,\ 120,\ 3\times 10^{-3})$,  (c) $(\Omega, \Delta_{2-}, g_z, \omega_m, \gamma) = (0.20,\ 0,\ 1.0,\ 120,\ 3\times 10^{-3})$. (d) $(\Omega, \Delta_{2-}, g_x, \omega_m, \gamma) = (0.20,\ 0,\ 1.0,\ 120,\ 3\times 10^{-3})$, (e) $(\Omega, \Delta_{2-}, g_x, g_z, \gamma) = (0.20,\ 0,\ 1.0,\ 1.0,\ 3\times 10^{-3})$, and (f) $(\Omega, \Delta_{2-}, g_x, g_z, \omega_m) = (0.20,\ 0,\ 1.0,\ 1.0,\ 120)$. All frequencies, couplings, and rates are expressed in units of $\kappa$, and the mechanical bath is assumed to be cold, $n_\mathrm{th} = 0$.
    }
    \label{fig:fig3}
\end{figure*}

\section{Comparison of Full and Effective Master Equations}\label{Appendix:C}

We benchmark the reduced master equation~\eqref{eq:EffME} against the full driven qubit–mechanics model~\eqref{eq:master} in Fig.~\ref{fig:fig2}. For both simulations we use the same parameters:
\(g_z/2\pi=6~\mathrm{MHz}\), \(g_x=0.1\,g_z\), \(\omega_m/2\pi=100~\mathrm{MHz}\), \(\kappa/2\pi=100~\mathrm{kHz}\), \(\varepsilon=2g\), and \(\gamma/2\pi=15~\mathrm{Hz}\),
with the qubit driven at \(\omega_d=\omega_q=2\omega_m\) and a zero-temperature mechanical bath (\(n_\mathrm{th}=0\)). The mechanical Hilbert space is truncated to \(N=50\).
The effective model uses the rates and coherent terms defined below Eq.~\eqref{eq:EffME} (in particular \(\Gamma_{1,2}^{\pm}\), \(\delta_k\), and \(\chi\)),
while the full model~\eqref{eq:master} evolves the combined qubit–mechanics density matrix with the explicit time-dependent Hamiltonian and the qubit damping \(\kappa\). For the numerical simulations, we transform the full master equation [Eq.~\eqref{eq:master}] to the frame rotating at the mechanical frequency \(\omega_m\), and the reduced master equation [Eq.~\eqref{eq:EffME}] to the frame rotating at the effective mechanical frequency \(\omega_m^{\mathrm{eff}}\). This facilitates a direct comparison between Wigner functions at different times by removing the overall rotations in phase space.

Fig.~\ref{fig:fig2} quantifies the agreement between the full model~\eqref{eq:master} and reduced effective model~\eqref{eq:EffME}. In the right column, we plot the Uhlmann fidelity between the reduced mechanical state from the full model and that from the effective model. For two density operators \( \rho \) and \( \sigma \), the fidelity is defined as  
\begin{equation}
F(\rho,\sigma) = \left[\operatorname{Tr}\!\left(\sqrt{\sqrt{\rho}\,\sigma\,\sqrt{\rho}}\right)\right]^2,
\label{eq:fidelity}
\end{equation}
which takes values in \([0,1]\) and reduces to the squared overlap for pure states. In our case,  
\(\rho_{\mathrm{full},m}(t) = \operatorname{Tr}_{q}[\rho_{\mathrm{full}}(t)]\) and \(\rho_{\mathrm{eff}}(t) = \bar{\rho}_m(t)\). The fidelity (top right panel) rises rapidly after the initial transient, surpasses \(0.95\) by \(\kappa t \simeq 15\), and then saturates near \(0.98\) with small residual ripples. The mean phonon number (bottom right) from the effective model (red dashed) tracks the full result (black solid) throughout: after the rapid ramp it remains within a few percent and is slightly lower than the full value at long times. The left panels compare Wigner snapshots at matched times \(\kappa t=0,\,1.31,\,19.66,\,39.00\): both descriptions exhibit the initial vacuum peak, the bifurcation into two lobes, the development of parity-protected interference fringes, and the final two-lobe steady state. The small early-time deviations and the steady-state offset are consistent with the approximations used to derive Eq.~(\ref{eq:EffME}) (e.g., neglect of counter-rotating terms and the assumption of fast qubit relaxation). Together, these benchmarks support the validity of the reduced master equation in the regime \(\kappa \gg \{\Gamma_2^-,\,\Gamma_1^\pm,\,\gamma\}\) and \(\omega_q \simeq \omega_d \simeq 2\omega_m\).

To assess the robustness of the effective master equation beyond the two operating points shown in Figs.~\ref{fig:fig1} and \ref{fig:fig2}, we carry out a systematic parameter scan summarized in Fig.~\ref{fig:fig3}. The derivation of Eq.~(\ref{eq:EffME}) assumes weak transverse and longitudinal couplings \(g_{x,z} \ll \omega_m\), a strongly damped qubit with \(\kappa \gg \{\Gamma_2^{-},\Gamma_1^{\pm},\gamma\}\), operation close to the two-phonon resonance \(\Delta_{2-}  \simeq 0\). In Fig.~\ref{fig:fig3}(a)–(f), we then vary one control parameter at a time—the drive amplitude \(\Omega/\kappa\), the two-phonon detuning \(\Delta_{2-}/\kappa\), the couplings \(g_x/\kappa\) and \(g_z/\kappa\), the mechanical frequency \(\omega_m/\kappa\), and the mechanical damping \(\gamma/\kappa\)—while keeping all other parameters within this hierarchy. Across all scans, the fidelity between the effective and full dynamics remains very close to unity, typically above \(0.995\) and never dropping below about \(0.985\), confirming that the reduced description is quantitatively accurate over a broad and experimentally realistic range. The small oscillations visible in panel (b) simply track the variation of the qubit response functions with detuning and sideband position and do not signal any qualitative breakdown of the approximations.Noticeable deviations only begin to appear when we deliberately approach the boundary of the assumed regime, for example at larger transverse coupling in panel (c), where the perturbative treatment in \(g_x/\omega_m\) becomes less accurate. A similar trend is observed in panel (d) for larger longitudinal coupling, where the expansion in \(g_z/\omega_m\) becomes less controlled. In panel (e), increasing \(\omega_m/\kappa\) drives the system deeper into the regime of validity of the effective model, leading to a gradual increase in fidelity that is consistent with the perturbative structure in \(g_x g_z/\omega_m\). Finally, in panel (f), noticeable deviations arise only when the mechanical damping \(\gamma\) is increased to the point where it becomes comparable to the engineered qubit-induced rates, thereby weakening the separation of time scales that underlies the adiabatic elimination. Overall, these scans delineate the domain of validity of Eq.~(\ref{eq:EffME}) and show that the working points used in Figs.~\ref{fig:fig1} and \ref{fig:fig2} lie well inside a parameter region where the two-step effective theory faithfully reproduces the full dynamics.

\begin{figure*}[t]
    \centering
    \includegraphics[width=0.24\textwidth]{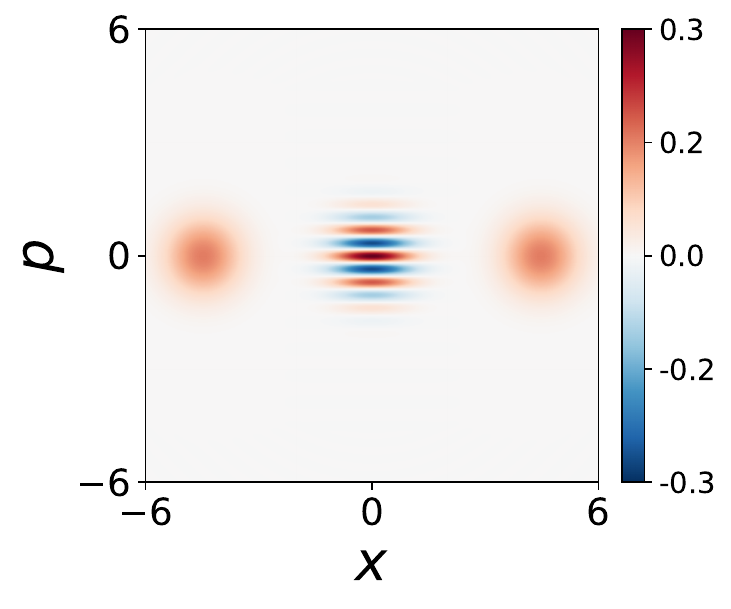}
    \includegraphics[width=0.24\textwidth]{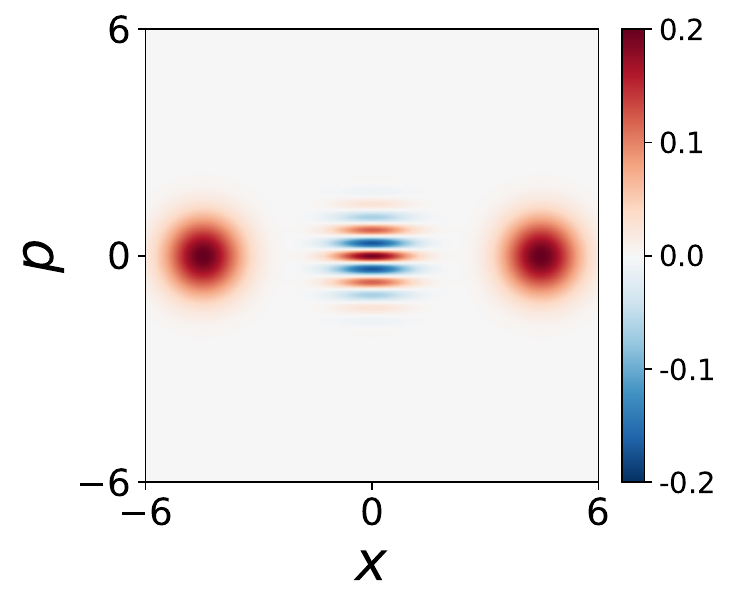}
    \includegraphics[width=0.24\textwidth]{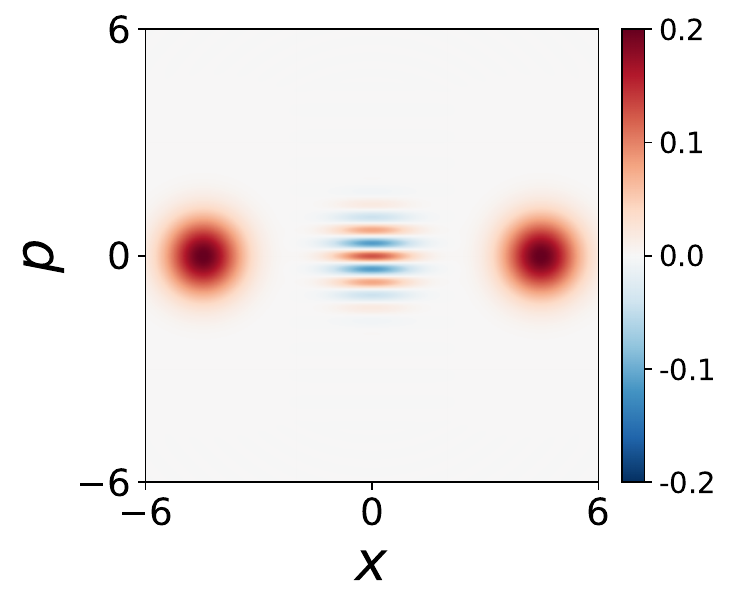}
    \includegraphics[width=0.24\textwidth]{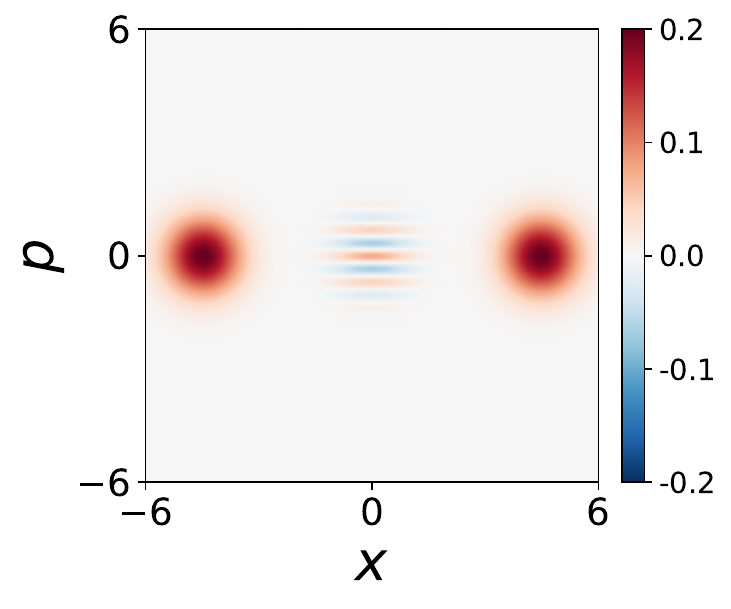}
    \caption{Wigner distributions \(W(x,p)\) of the mechanical resonator for different thermal occupations of its phonon bath. The panels, from left to right, correspond to \(n_{\mathrm{th}} = 0\), \(0.5\), \(1\), and \(2\), respectively, and show the Wigner function at a fixed evolution time \(\Gamma_{2}^{-} t = 3\). For \(n_{\mathrm{th}} = 0\), the steady state is a nearly ideal even Schr\"odinger-cat, characterized by two well-separated phase-space lobes and pronounced interference fringes. As \(n_{\mathrm{th}}\) increases, thermal phonons progressively smear out the interference pattern and reduce the negativity of \(W(x,p)\), illustrating the gradual degradation of cat-like nonclassicality under realistic mechanical decoherence. The system parameters are the same as in Fig.~\ref{fig:fig1}, except $\varepsilon = 10 g$. The evolution is governed by the effective master equation~(\ref{eq:EffME}) and the mechanical Hilbert space truncated at \(N=60\).}
    \label{fig:fig4}
\end{figure*}

{\color{black}
\section{Effects of thermal decoherence, Kerr nonlinearity, and sideband resolution}
\label{app:D}

In this appendix, we investigate how three non-ideal features of a realistic mechanical setup modify the cat-state physics described in Sec.~\ref{sec:results}. First, in Sec.~\ref{subsec:thermal}, we examine the impact of a finite-temperature mechanical bath and its associated single-phonon damping on the formation and robustness of the cat state. Second, in Sec.~\ref{subsec:kerr}, we analyze the role of the induced Kerr nonlinearity in the reduced mechanical Hamiltonian, clarifying when it can be neglected and how, in other regimes, it can either assist or hinder cat-state formation. Finally, in Sec.~\ref{subsec:sideband}, we relax the sideband-resolved assumption that $\omega_{m}\gg\kappa$ and study the resulting sideband-unresolved dynamics. In this regime, additional off-resonant and single-phonon processes weaken the effective two-phonon reservoir and diminish the Wigner negativity of the mechanical cat state.

\subsection{Thermal decoherence of the mechanical cat state}\label{subsec:thermal}

We begin by assessing how a finite-temperature phonon bath affects the dynamics of the mechanical resonator and the resulting cat state. In the effective master equation~(\ref{eq:EffME}), the mechanical environment enters through single-phonon dissipators with rates \(\gamma_{\pm}\), which grow with the thermal occupation \(n_{\mathrm{th}}\). These single-phonon processes compete directly with the engineered two-phonon channels characterized by \(\Gamma_{2}^{\pm}\). In the ideal limit where the intrinsic single-phonon damping is negligible compared to the engineered two-phonon rates and the bath occupation is very small, phonon-number parity is effectively conserved. In this regime, the two-phonon dissipator drives an initially even (odd) state into an even (odd) cat state. As \(n_{\mathrm{th}}\) increases, however, single-phonon jumps increasingly induce parity mixing and gradual thermalization. This mixing washes out quantum interference and progressively reduces the Wigner negativity of the mechanical steady state.

Figure~\ref{fig:fig4} illustrates this competition by showing  Wigner function \(W(x,p)\) for four bath occupancies, \(n_{\mathrm{th}} = 0, 0.5, 1,\) and \(2\) (panels from left to right), at a fixed evolution time \(\Gamma_{2}^{-} t = 3\). For \(n_{\mathrm{th}} = 0\), the steady state is a nearly ideal even Schr\"odinger-cat: two well-separated lobes in phase space are connected by a clear pattern of interference fringes with pronounced negative regions of \(W(x,p)\). At \(n_{\mathrm{th}} = 0.5\), the two lobes remain visible, but the fringes are already noticeably blurred, and the Wigner negativity is reduced, signalling a partial loss of coherence between the two components. For \(n_{\mathrm{th}} = 1\), the interference pattern is strongly suppressed and the phase-space distribution becomes broader and more classical-looking, although remnants of a bimodal structure can still be discerned. By \(n_{\mathrm{th}} = 2\), the Wigner function is essentially positive and closely resembles a displaced, thermally broadened state, indicating that the cat-like nonclassicality has been almost completely destroyed.

These results make explicit that, for a given engineered two-phonon rate \(\Gamma_{2}^{-}\), there is a practical upper bound on the tolerable thermal occupation. To sustain a mechanically generated cat state with visible interference fringes, the two-phonon timescale must remain much faster than the single-phonon thermalization timescale, roughly \(\Gamma_{2}^{-} \gg \gamma \left( 2n_{\mathrm{th}} + 1 \right)\). Figure~\ref{fig:fig4} thus provides a concrete, visual complement to the design rule that robust mechanical cat states require not only a strong engineered two-phonon reservoir, but also a sufficiently cold and weakly damped phonon environment.

\subsection{Interplay between engineered two-phonon dissipation and Kerr nonlinearity}\label{subsec:kerr}

In the effective master equation~(\ref{eq:EffME}), the mechanical mode is subject to two distinct nonlinear channels generated by the same underlying qubit--resonator couplings: 
(i) a dissipative two-phonon reservoir with cooling rate \(\Gamma_{2}^{-}\), and  (ii) a coherent Kerr nonlinearity characterized by the coefficient \(\delta_{k}\). In this subsection, we investigate both contributions and show that their relative importance is set by a simple dimensionless ratio. This provides a clear criterion for distinguishing reservoir-dominated and Kerr-dominated operating regimes for a fixed underlying Hamiltonian. To place our analysis in context, it is useful to contrast it with earlier two-photon cat-state schemes. In Refs.~\cite{PhysRevA.99.022302,PhysRevA.110.013711}, the authors derive an effective qubit--resonator Hamiltonian in the two-photon (or magnon) regime and then introduce dissipation at the master-equation level. As a consequence, the coherent Kerr nonlinearity generated by the same virtual processes does not play an explicit, separate role in their description of the cat-state dynamics. In our treatment, by contrast, this Kerr term is kept throughout and analyzed on the same footing as the engineered two-phonon dissipation.

Following the adiabatic elimination in Appendix~\ref{Appendix:B}, the two-phonon processes are captured by the qubit response function \(S_{2\pm}(\Delta)\), given in Eq.~\eqref{eq:S2pm} and evaluated at the detunings \(\Delta_{2\pm}\).
At the two-phonon resonance \(\Delta_{2-} = \omega_q - 2\omega_m = 0\), and in the sideband-resolved limit \(\omega_m \gg \kappa\), Eq.~\eqref{eq:S2pm} yields
\begin{equation}
S_{2-}(0) = \frac{2}{\kappa},
\qquad
S_{2+}(4\omega_m)
\simeq \frac{\kappa}{32\omega_m^2}
- i\,\frac{1}{4\omega_m}.
\label{eq:D_S2_eval}
\end{equation}
Inserting Eq.~\eqref{eq:D_S2_eval} into Eqs.~\eqref{eq:delta_k}--\eqref{eq:Gamma_pm}, and noting that
$\mathrm{Im}\,S_{2-}(0)=0$ at the two-phonon resonance, gives
\begin{align}
\Gamma_2^{-}  = \frac{4 g^2}{\kappa},
\qquad\qquad
\delta_k \simeq -\,\frac{g^2}{4\omega_m}.
\label{eq:D_K_result}
\end{align}
The ratio of coherent to dissipative nonlinearities then follows as
\begin{equation}
\frac{\delta_k}{\Gamma_2^{-}}
\simeq -\,\frac{\kappa}{16\omega_m}.
\label{eq:D_K_over_Gamma2_raw}
\end{equation}
Deep in the sideband-resolved regime \(\omega_m \gg \kappa\), we therefore have \(|\delta_k|/\Gamma_2^{-} \ll 1\), and the dynamics are dominated by the engineered two-phonon reservoir.
Nevertheless, the Kerr term still ``attempts'' to generate a cat state whose amplitude is parametrically larger than that of the reservoir-engineered two-phonon cat state~\cite{PhysRevLett.130.213604}.
As \(\omega_m/\kappa\) is reduced, the Kerr channel becomes increasingly important and eventually competes with, or even dominates over, the dissipative channel.

\subsection{Unresolved-sideband regime and breakdown of cat-state stabilization}\label{subsec:sideband}

The cat-state stabilization results presented in Fig. \ref{fig:fig1} are computed under the sideband-resolved condition \(\kappa \ll \omega_m\), which ensures that the two-phonon sideband at \(\omega_q \simeq 2\omega_m\) is spectrally well separated. It is therefore natural to ask how this picture is modified when the qubit linewidth becomes comparable to the mechanical frequency and the system enters the unresolved-sideband regime, \(\kappa \gtrsim \omega_m\).

The competition between engineered two-phonon cooling and qubit-induced single-phonon processes can be quantified in terms of the response functions \(S_{\pm}\) and \(S_{2\pm}\) defined in Eq.~\eqref{eq:S_pm}. At the two-phonon resonance \(\Delta_{2-} = 0\), one has \(\mathrm{Re}\,S_{2-} = 2/\kappa\), whereas for \(\omega_q \simeq 2\omega_m\) the single-phonon response at the lower sideband is
\begin{equation}
\mathrm{Re}\,S_{-} = \frac{\kappa/2}{(\kappa/2)^2 + \omega_m^2}.
\end{equation}
From Eq.~\eqref{eq:Gamma_pm}, the ratio of the induced one-phonon cooling rate to the two-phonon cooling rate is therefore
\begin{equation}
\frac{\Gamma_1^-}{\Gamma_2^-} = \frac{g_x^2}{g^2}\, \frac{\mathrm{Re}\,S_{-}}{\mathrm{Re}\,S_{2-}} \simeq \frac{g_x^2}{g^2}\, \frac{\kappa^2}{4\big[(\kappa/2)^2 + \omega_m^2\big]}.
\label{eq:coolingratio}
\end{equation}
In the sideband-resolved limit \(\kappa \ll \omega_m\) this ratio is suppressed as
\begin{equation}
\frac{\Gamma_1^-}{\Gamma_2^-} \propto \left(\frac{\kappa}{\omega_m}\right)^2
\end{equation}
(up to the fixed prefactor \(g_x^2/4g^2\)), so the engineered two-phonon channel dominates and phonon-number parity is approximately conserved. By contrast, once \(\kappa\) becomes comparable to \(\omega_m\), the dimensionless combination
\begin{equation}
\frac{\kappa^2}{4\big[(\kappa/2)^2 + \omega_m^2\big]}
\end{equation}
appearing in Eq.~\eqref{eq:coolingratio} is of order unity, and \(\Gamma_1^-\) and \(\Gamma_2^-\) become similar in magnitude. At the same time, off-resonant contributions to \(\Gamma_1^+\), \(\Gamma_2^+\), and the Lamb and Kerr shifts are no longer parametrically small. The hierarchy required for robust cat-state stabilization, \(\Gamma_2^- \gg \{\Gamma_1^\pm, \Gamma_2^+, \gamma_\pm\}\), is then violated. 
In the unresolved-sideband regime the engineered two-phonon reservoir can therefore no longer maintain a clean parity-protected steady state. Frequent single-phonon jumps rapidly mix even and odd manifolds, and the long-time state approaches a thermally broadened squeezed state rather than a genuine Schr\"odinger-cat. Weak cat-like signatures, such as a distorted bimodal structure with small Wigner negativity, may survive for \(\kappa/\omega_m \lesssim 1\) and suitably strong couplings. However, large, well-resolved mechanical cat states require operation in the sideband-resolved regime where the two-phonon sideband is spectrally isolated.

\section{Two-phonon cooling and cat state stabilization}
\label{app:E}

In this appendix, we derive an analytic estimate for the steady-state mean phonon number of the mechanical resonator in the cat-state regime. Starting from the full effective master equation \eqref{eq:EffME}, we identify the dominant two-phonon processes near the second-order sideband and exploit the resulting engineered two-phonon reservoir to construct a simple cat-state--based approximation. This makes explicit that the same two-phonon dynamics that stabilize the superposition also act as an efficient cooling mechanism for the resonator.
Throughout, we focus on the parameter regime used in Sec.~\ref{sec:results}: the drive is tuned close to the qubit, \(\omega_d \simeq \omega_q\), the two-phonon resonance is nearly satisfied, \(\Delta_{2-}\simeq 0\). In addition, we work in the sideband-resolved limit \(\omega_m \gg \kappa\) so that the two-phonon sideband is spectrally well isolated. In this regime, the engineered two-phonon exchange both generates the cat state and dominates the energy flow, and the following hierarchy is satisfied:

\begin{figure}[t]
    \centering
    \includegraphics[width=0.5\columnwidth]{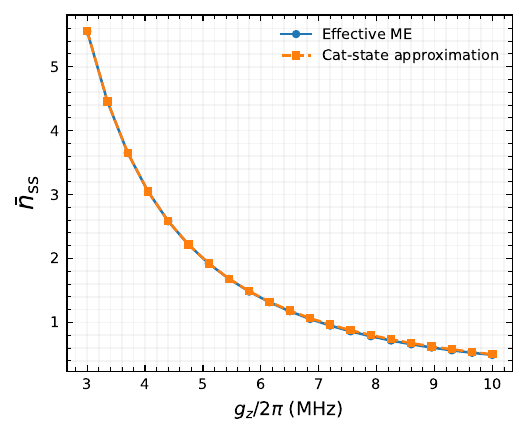}
    \caption{Steady-state mean phonon number $\bar{n}_{\mathrm{ss}}$ as a function of the longitudinal coupling $g_z/2\pi$.  The blue solid line with circles shows the result obtained from the effective master equation \eqref{eq:EffME}, while the orange dashed line with squares corresponds to the cat-state approximation given in Eq. \eqref{eq:A12}. Parameters: $g_x/g_z = 0.1$, $\omega_m/2\pi = 100~\mathrm{MHz}$, $\omega_q = 2\omega_m$, $\kappa/2\pi = 100~\mathrm{kHz}$, $\gamma/2\pi = 15~\mathrm{Hz}$, $n_{\rm th} = 5$, $\Omega/2\pi = 0.10~\mathrm{MHz}$, and $N = 80$.}
    \label{fig:fig5}
\end{figure}

\begin{enumerate}
\item Two-phonon cooling dominates: $\Gamma_2^- \gg \{\Gamma_2^+,\, \Gamma_1^\pm,\, \gamma_\pm\}.$
This reflects the enhancement of \(\mathrm{Re}\,S_{2-}\) at \(\Delta_{2-}=0\) and the suppression of the off-resonant channels \(\Gamma_2^+\) and \(\Gamma_1^\pm\) (cf.\ Eqs.~(\ref{eq:Gamma_pm}) and~(\ref{eq:S_pm})).

\item Cold mechanical bath: the intrinsic environment of the resonator is effectively at sufficiently low temperature on the scale of \(\omega_m\), so that \(\gamma_+ \ll \{\gamma_-, \, \Gamma_2^-\}\).

\item Moderate Kerr nonlinearity: the Kerr rate \(|\delta_k|\) generated by the same virtual processes is smaller than \(\Gamma_2^-\) in the sideband-resolved limit (Appendix~\ref{Appendix:B}).
\end{enumerate}
Within this hierarchy, we retain only the coherent two-phonon drive and the dominant two-phonon cooling channel, the master equation \eqref{eq:EffME} then reduces to
\begin{equation}
\dot{\bar\rho}_m = -\,i\left[\tilde\omega_m a^\dagger a +\chi a^2 +\chi^* a^{\dagger 2}, \ \bar\rho_m\right] +\Gamma_2^-\,\mathcal{D}[a^2]\bar\rho_m,
\label{eq:A3}
\end{equation}
here, $\tilde{\omega}_m \equiv \omega_m + \delta_1 + \delta_2$. Eq.~(\ref{eq:A3}) is a standard two-phonon squeezing-plus-loss model: the coherent term pumps phonon pairs into the resonator, while the dissipator removes energy in pairs and preserves phonon-number parity. The steady state of this equation is well described by a Schr\"odinger-cat state stabilized by an engineered two-phonon reservoir \cite{Mirrahimi_2014}. The Liouvillian corresponding to Eq.~(\ref{eq:A3}) can be rewritten, in the frame rotating at frequency $\tilde{\omega}_{m}$, and in the standard reservoir-engineering form \cite{Leghtas2015}
\begin{equation}
\mathcal{L}\bar\rho_m \simeq \Gamma_2^-\,\mathcal{D}[a^2-\alpha_*^2]\bar\rho_m,
\label{eq:A7}
\end{equation}
provided we choose the complex parameter \(\alpha_*^2\) such that
\begin{equation}
\alpha_*^2 = \frac{2\chi}{\Gamma_2^-},
\label{eq:A8}
\end{equation}
In this representation, the dissipator in Eq. \eqref{eq:A7} drives the oscillator into the dark subspace of the jump operator \(a^2-\alpha_*^2\), defined by
\begin{equation}
(a^2-\alpha_*^2)\ket{\psi_{\rm dark}}=0.
\end{equation}
The dark manifold is spanned by the even and odd Schr\"odinger-cat states built from the coherent states \(\ket{\pm\alpha_*}\),
\begin{equation}
\ket{\mathcal{C}_\pm} \propto \ket{\alpha_*}\pm\ket{-\alpha_*},
\label{eq:A9}
\end{equation}
which are both eigenstates of \(a^2\) with eigenvalue \(\alpha_*^2\). Because the engineered dynamics preserve phonon-number parity and the resonator is initialized in an even state, the evolution selects the even cat \(\ket{\mathcal{C}_+}\), while weak residual processes lift the degeneracy between \(\ket{\mathcal{C}_+}\) and \(\ket{\mathcal{C}_-}\). In the following we use this cat-state ansatz to estimate the steady-state phonon number.

The mean phonon number of the even cat state \(\ket{\mathcal{C}_+}\) built from amplitude \(\alpha_*\) is given by
\begin{equation}
n_{\rm ss} \simeq \langle \hat n \rangle_{\mathcal{C}_+} = |\alpha_*|^2 \tanh |\alpha_*|^2.
\label{eq:A10}
\end{equation}
Combining Eq.~(\ref{eq:A10}) with Eq.~(\ref{eq:A8}) gives our cat-state approximation for the steady-state occupancy in terms of the engineered two-phonon parameters,
\begin{equation}
n_{\rm ss} \simeq \frac{2|\chi|}{\Gamma_2^-}\, \tanh\!\left(\frac{2|\chi|}{\Gamma_2^-}\right).
\label{eq:A11}
\end{equation}
In the cat-state regime realized in Sec. \ref{sec:results}, the engineered amplitude satisfies \(|\alpha_*|^2 = 2|\chi|/\Gamma_2^- \gtrsim 3\), so that \(\tanh |\alpha_*|^2 \approx 1\). In this limit Eq.~(\ref{eq:A11}) simplifies to the particularly transparent expression
\begin{equation}
n_{\rm ss} \simeq \frac{2|\chi|}{\Gamma_2^-}, \qquad (|\alpha_*|^2 \gg 1).
\label{eq:A12}
\end{equation}
Using Eqs.~(\ref{eq:S_pm})--(\ref{eq:chi}) at \(\Delta_{2-}=0\), Eq.~(\ref{eq:A12}) can also be written as \(n_{\rm ss} \simeq \Omega\omega_m/(2g_xg_z)\). Eqs.~(\ref{eq:A11})--(\ref{eq:A12}) make explicit that the steady-state phonon number is set primarily by the ratio of the coherent two-phonon pump \(|\chi|\) to the engineered two-phonon cooling rate \(\Gamma_2^-\).  

Figure~\ref{fig:fig5} illustrates how the steady-state mean phonon number \(\bar{n}_{\rm ss}\) depends on the longitudinal coupling strength \(g_z\). The solid curve shows \(\bar{n}_{\rm ss}\) obtained from the full effective master equation~\eqref{eq:EffME}, while the dashed curve displays the cat-state approximation given in Eq. \eqref{eq:A12}. Over the scanned range \(g_z/2\pi \in [3,10]~\mathrm{MHz}\), both curves exhibit a clear reduction of \(\bar{n}_{\rm ss}\) with increasing \(g_z\), demonstrating enhanced two-phonon cooling relative to the bare thermal occupation \(n_{\rm th}=5\). The close agreement between the two curves confirms that the simple cat-state approximation captures both the qualitative trend and the quantitative scale of the cooling. Figure~\ref{fig:fig5} shows that, in the parameter regime where the mechanical cat state is realized, the engineered two-phonon reservoir significantly reduces the resonator occupation below the bare thermal level. Together with the reservoir-engineering picture above, this suggests that the resonator is cooled towards the cat-state manifold rather than towards a simple thermal state.
}

\end{widetext}

\bibliography{CatState.bib}

\end{document}